\documentclass[aps,prl,twocolumn,amsmath,amssymb,nofootinbib,superscriptaddress,longbibliography]{revtex4-2}

\usepackage{graphicx}
\usepackage{dcolumn}
\usepackage{bm}

\usepackage{xcolor}
\usepackage{comment}

\usepackage[colorlinks=true,bookmarks=false,allcolors=blue]{hyperref} 

\newcommand{\fla}[1]{\begin{flalign}#1\end{flalign}}

\begin{document}

\title{Exceptional swallowtail degeneracies in driven-dissipative quadrature squeezing}

\author{Polina Blinova}
\thanks{These authors contributed equally to this work.}
\affiliation{Department of Physics, McGill University, 3600 rue University, Montreal, QC, H3A 2T8, Canada.}
\affiliation{School of Applied and Engineering Physics, Cornell University, Ithaca, NY, 14850, USA.}

\author{Evgeny Moiseev}
\thanks{These authors contributed equally to this work.}
\affiliation{Department of Physics, McGill University, 3600 rue University, Montreal, QC, H3A 2T8, Canada.}

\author{Kai Wang}
\email{k.wang@mcgill.ca}
\affiliation{Department of Physics, McGill University, 3600 rue University, Montreal, QC, H3A 2T8, Canada.}

\date{\today}

\begin{abstract}
We show that swallowtail catastrophe consisting of various-order non-Hermitian and Hermitian degeneracies naturally exists in the dynamics of two-mode driven-dissipative quadrature squeezing systems that break pseudo-Hermiciticy by judiciously engineered losses. We reveal that the swallowtail degeneracy structure enables nontrivial braiding of complex eigenvalues by looping around an exceptional line in the parameter space. Our findings provide a comprehensive understanding of the degeneracy geometry in two-mode driven-dissipative bosonic quadratic systems, opening new pathways toward topologically nontrivial control of Gaussian states.

\end{abstract}

\maketitle


Swallowtail catastrophe is one of the elementary catastrophes in Arnold’s ADE classification \cite{arnol2003catastrophe} that appears in different branches of physics from magneto-hydrodynamics \cite{arnoldMHD83} to attosecond generation \cite{Raz2012}. Of particular research interest and practical relevance in applying catastrophe theory is the analysis of the degeneracy structure of ubiquitous open systems characterized by non-Hermitian Hamiltonians or dynamical matrices. In non-Hermitian systems, it has been reported that many exotic effects originate from the intricate degeneracy morphology associated with exceptional points (EPs)~\cite{ding2022non,heiss2012physics,ozdemir2019parity,parto2020non}.  
Systems exhibiting swallowtail degeneracies, composed of non-Hermitian degeneracies of different orders and occasionally Hermitian degeneracies, present a richer degeneracy morphology beyond simple EPs and have recently attracted research interest.
Very recent efforts revealed that swallowtail degeneracy naturally exists classical non-Hermitian systems~\cite{Hu2023}. However, so far, swallowtail degeneracy has only been found in classical non-Hermitian systems, and it remains an open question whether swallowtail degeneracy structures exist in quantum systems.

Bosonic quadratic Hamiltonians constitute an important class of quantum systems associated with Gaussian states, which underpin many effects ranging from the behaviors of magnons \cite{KatsuraPhysRevLett.104.066403,tserkovnyak,owerre2016first} to the quadrature squeezing of light \cite{walls1983squeezed,schumakerPhysRevA.31.3068,andersen201630}. In particular, in the research field of optics and photonics, squeezed states of light have showcased many powerful applications, from quantum sensing for gravitational wave detection \cite{caves,aasi2013enhanced} to quantum computation \cite{braunsteinRevModPhys.77.513} and communication \cite{shapiro1055958}. The dynamics of bosonic quadratic systems are characterized by non-Hermitian dynamical matrices, even if the Hamiltonian is Hermitian \cite{wang2019non,PhysRevX.8.041031,flynn2020deconstructing,Roy:21,ShengwangPhysRevLett.128.173602}. Realistic quadrature squeezing systems additionally have dissipation; for example, in photonic structures, losses are inevitable, leading to more intricate degeneracy structures and associated dynamics. 
Understanding such complex degeneracy morphology is of great importance as squeezing is an essential ingredient in the universal control of Gaussian states~\cite{Braunstein,adesso2014continuous}.

Here, we show that swallowtail degeneracies exist in the dynamics of driven-dissipative quadrature squeezing systems by a thorough analysis of the dynamical matrices of two-mode bosonic quadratic systems. More specifically, we perform detailed symmetry analysis and reveal the enabling role of breaking pseudo-Hermiticity via judiciously added losses in observing swallowtail degeneracies in such two-mode squeezing dynamics. 
Moreover, we will show the implication of identifying the swallowtail degeneracy with an example of forming a nontrivial braid with the complex eigenvalue of the driven-dissipative squeezing system when the control parameters go along topologically nontrivial loops without encountering any degeneracies.

\begin{figure}[!t]
	\centering
	\includegraphics[width=1\linewidth]{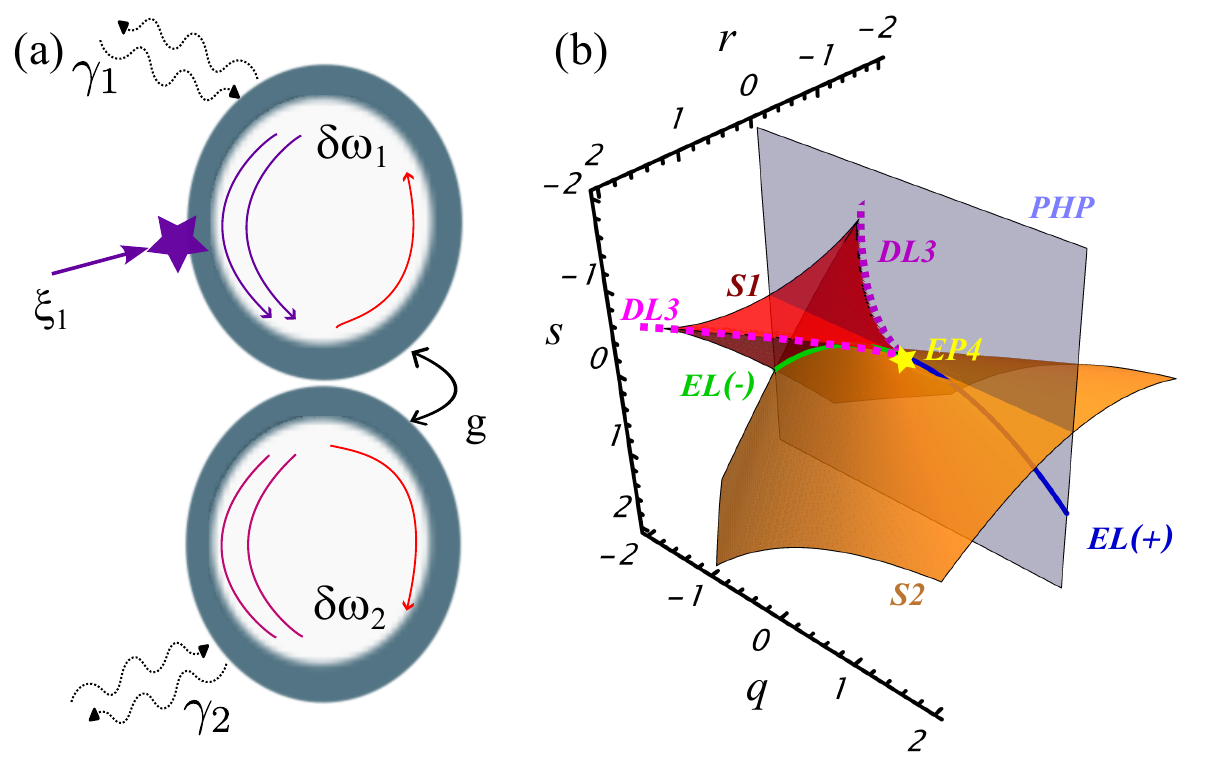}
	\caption{Swallowtail degeneracy in the dynamics of two-mode driven-dissipative squeezing system. (a) Two coupled optical ring-resonators with resonant frequencies $\delta \omega_{1,2}$. The interplay among squeeze-type driving $\xi_{1}$, beam-splitter-type coupling $g$, and losses $\gamma_{1,2}$ gives rise to the swallowtail degeneracy. (b) Swallowtail degeneracy in the control parameter space spanned by three real parameters $q, r, s$. The red/orange surface is formed by exceptional and diabolical points. Highlighted on the surface includes a fourth-order EP (yellow star), an exceptional line EL($-$) at the intersection of subsurface S1 and S2 (green line) with two pairs of degenerate eigenvalues, two diabolic lines (DL3) with three-fold Hermitian degeneracies (pink dashed lines). Moreover, there is an outgoing two-fold exceptional line EL($+$) with two pairs of degenerate eigenvalues (blue line). The light-blue-shaded plane is not a part of the swallowtail degeneracy, which is shown to indicate the pseudo-Hermitian plane (PHP) subspace of the $\mathbb{R}^3$ parameter space. }
	\label{main}
\end{figure}

While our findings are general to arbitrary bosonic quadratic Hamiltonians with two modes over complex coefficients (Supplemental Sec. I~\cite{supp}), for the sake of argument, here we restrict our attention to the following specific Hamiltonian: 
\fla{
	\hat{\mathcal{H}} &=
	\delta\omega_1 \hat{a}^{\dagger}_1 \hat{a}_1 + \delta\omega_2 \hat{a}^{\dagger}_2 \hat{a}_2 \nonumber \\ &+ i g (  \hat{a}^{\dagger}_1 \hat{a}_2 -  \hat{a}_1 \hat{a}^{\dagger}_2  )  
	+ \frac{i \xi_1 }{2}\left( \hat{a}^{\dagger}_1 \hat{a}^{\dagger}_1 -  \hat{a}_1 \hat{a}_1  \right), 
	\label{Ham}
}
where $\hat{a}_{1(2)}$ $(\hat{a}^{\dagger}_{1(2)})$ is the annihilation (creation) operator for mode 1(2). Such a two-mode system can be illustratively represented as two coupled optical ring resonators shown in Fig. \hyperref[main]{1(a)}, where mode 1(2) has a resonance frequency of $\delta\omega_{1(2)} \in \mathbb{R}$ defined with respect to a properly chosen reference frame. The two modes experience a linear coupling with rate $g\in \mathbb{R}$, which is a beam-splitter-type interaction. Mode 1 experiences a parametric gain (squeezing) with a real-valued parameter $\xi_{1}$. Such a parametric driving can be achieved, e.g., by nonlinear optics processes with undepleted optical pump(s).
In the presence of losses at an input-output port for each mode with rates $\gamma_{1,2}$ for each mode, the dynamics of the system are governed by linear Heisenberg-Langevin equations of the form\nocite{Walls2008,michael_spivak_comprehensive_1979,melkani,vassiliev1995topology,batorova2013desingularization,spivak2018calculus,ARNON198837} $\partial_t\mathbf{a}  = \tilde{\mathcal{E}} \mathbf{a} + \mathcal{K}\mathbf{a}_{\text{in}}\label{eq::EOM}
$~\cite{Walls2008}, where $\textbf{a} = (\hat{a}_1,\hat{a}_2,\hat{a}^\dag_1,\hat{a}^\dag_2)^T$, and the fluctuating operator $\textbf{a}_{\text{in}}=(\hat{a}_{1,\mathrm{in}},\hat{a}_{2,\mathrm{in}},\hat{a}^\dag_{1,\mathrm{in}},\hat{a}^\dag_{2,\mathrm{in}})^T$ is associated with a diagonal matrix incorporating the loss rates $\mathcal{K}=\text{diag}(\sqrt{\gamma_1},\sqrt{\gamma_2},{\sqrt{\gamma_1}},\sqrt{\gamma_2})$ according to the dissipation-fluctuation theorem \cite{Gardiner1985}.The explicit form of the dynamical matrix $\tilde{\mathcal{E}}$ for this system is
\fla{{\footnotesize
		\tilde{\mathcal{E}}=\begin{pmatrix}
			-\gamma_1/2 -i\delta \omega_1 & g & \xi_1 & 0 \\
			-g & - \gamma_2/2 - i\delta \omega_2 & 0  &0 \\
			\xi_1 & 0& -\gamma_1/2 + i\delta \omega_1 &  g \\
			0 & 0 & -g & -\gamma_2/2 + i \delta \omega_2 
	\end{pmatrix}}.\label{dyn}
}
This dynamical matrix $\tilde{\mathcal{E}}$, which is non-Hermitian, is of central importance to this system as it governs the dynamics. Without loss of generality, we base our analysis on a traceless version of the dynamical matrix $\mathcal{E} \equiv \tilde{\mathcal{E}} +\frac{\gamma_+}{4} I_4$  with $\gamma_{+} \equiv \gamma_1 + \gamma_2$ and $I_4$ being the $4\times 4$ identity matrix.  The degeneracy structure of the dynamical matrix is fully captured by the traceless matrix $\mathcal{E}$, since the eigenvectors are the same and the eigenvalues are simply shifted by $-\gamma_+/4$.

Now, we establish the relation between the symmetries of the dynamical matrix $\mathcal{E}$ and its characteristic polynomial as an important step to understanding its degeneracy morphology. It is known that bosonic quadratic Hamiltonians exhibit a mathematical analog of the particle-hole symmetry~\cite{phs}, which can be presented as
$\mathcal{E}^{*} = \tau_x \mathcal{E} \tau_x$, where $\tau_j=\sigma_j \otimes I_2$, with $\sigma_j$ ($j=x,y,z$) being a Pauli matrix, $I_2$ being the $2\times 2$ identity matrix, and $\otimes$ denoting the Kronecker (tensor) product. 
Most of the studied dynamical matrices of bosonic quadratic systems are closed systems, which additionally exhibit pseudo-Hermiticity~\cite{Mostafazadeh}, described by 
$\mathcal{E}^{\dagger} = -\tau_z \mathcal{E} \tau_z$. In our specific example of Eq.~\eqref{dyn}, pseudo-Hermiticity corresponds to the case where $\gamma_1=\gamma_2=0$. We find that for a general $\mathcal{E}$ of a two-mode bosonic quadratic system that exhibits both the particle-hole symmetry and the pseudo-Hermiticity, its characteristic polynomial is in the form of
\fla{p(\lambda) = \lambda^4 + q \lambda^2  + s,\label{chp2}}
where $q, s \in \mathbb{R}$. The control parameter space is thus $\mathbb{R}^2.$ Such a conclusion is not model-specific and solely comes from symmetry arguments of $\mathcal{E}$, where a proof can be found in Supplemental Sec. I B~\cite{supp}.

We find that the degeneracy structure is a swallowtail catastrophe if the pseudo-Hermiticity is nontrivially broken, e.g., by an uneven dissipation of the modes that couple to the output at different rates. In our specific example of Eq.~\eqref{dyn}, a sufficient condition that nontrivially breaks pseudo-Hermiticity is having $\gamma_1 \neq \gamma_2$, i.e., asymmetric losses for the two modes.
The connection of broken pseudo-Hermiticity and the swallowtail degeneracy structure can be seen by calculating the
characteristic polynomial of $\mathcal{E}$ that only exhibits particle-hole symmetry, which is in the form
\fla{p(\lambda) = \lambda^4 + q \lambda^2 +r \lambda + s,\label{chp}}
where $q, r, s \in \mathbb{R}$. 
Compared to Eq.~\eqref{chp2}, there is one more term of first order with a real coefficient $r$; hence the control parameter space becomes $\mathbb{R}^3$.
Likewise, such a conclusion does not depend on a specific model of two-mode bosonic quadratic Hamiltonians (see Supplemental Sec. I C~\cite{supp}).
Such a depressed quartic polynomial with real parameters precisely gives rise to the swallowtail catastrophe~\cite{arnold2006mathematical}.
Specifically, the eigenvalue degeneracies are the zero locus of the discriminant $D(q, r, s)$ of $p(\lambda)$ in Eq.~\eqref{chp} forms the "swallowtail" surface embedded in the $\mathbb{R}^3$ control parameter space [Fig.~\hyperref[main]{1(b)}], which hosts degeneracies of various orders and types. By contrast, without breaking pseudo-Hermiticity, the control parameter space is limited to an $\mathbb{R}^2$ subspace, indicated as the pseudo-Hermitian plane (PHP) in Fig.~\hyperref[main]{1(b)}.

While our finding of the swallowtail degeneracy structure is general for two-mode bosonic quadratic Hamiltonians, here we use our specific model (taking $\delta\omega_2=0$) to show an analytically solvable inverse map from the $\mathbb{R}^3$ control parameter space back to the Hamiltonian parameters. The map from the parameters of $\hat{\mathcal{H}}$ is defined by:
\fla{
	q=2\left(g^2-\gamma_{-}^2/16 \right)-u, \qquad
	r = \gamma_{-} u/2, \nonumber\\
	s = \left(g^2 - \gamma_{-}^2/16 \right)^2 - \gamma_{-}^2u/16,
	\label{eq::map}
}
where we have defined $u \equiv \xi_1^2-\delta\omega_1^2$ and $\gamma_{-} \equiv \gamma_1 - \gamma_2$. A sufficient condition for the map in Eq.~\eqref{eq::map} to have a local inverse at each point of parameter space is to ensure a non-vanishing determinant of the Jacobian of the transformation in Eq.~\eqref{eq::map} \cite[Theorem~2-11]{spivak2018calculus}. We formulate the invertibility conditions in a more general context in the Supplemental Sec. II A \cite{supp}.

In the following part of this Letter, we provide a detailed discussion on the eigenvalue and eigenvector degeneracies embedded in the swallowtail [Fig. \hyperref[main]{1(b)}] based on our specific model. 
The surface gives rise to both diabolical (Hermitian) and exceptional (non-Hermitian) degeneracies. In stark contrast to diabolical degeneracies, which are manifested by coalescing eigenvalues, an exceptional point (EP) is additionally accompanied by degenerate eigenstates. 
Here, the swallowtail is comprised of eigenvalue degeneracies of at least second order and is symmetric with respect to the PHP. It hosts a fourth-order degeneracy point (named EP4, denoted as the yellow star) where all the four eigenvalues coalesce. Depending on the Hamiltonian parameters, such a defective degeneracy can be either a fourth-order exceptional point where all four eigenvectors coalesce or a combination of a second-order exceptional and a second-order diabolical point where only two eigenvectors are degenerate (see Supplementary Sec. II C~\cite{supp}). 
The EP4 divides the exceptional line EL into two segments: EL($+$) that goes outside of surfaces S1 and S2, denoted by the blue line, hosts a pair of second-order defective degeneracies of imaginary eigenvalues; EL($-$) (green) at the intersections between surfaces S1 and S2 hosts a pair of second-order defective degeneracies of real eigenvalues. 
We also note the close connection of the degeneracies on the swallowtail with the existence of quantum nondemolition measurement (QND) operators, which goes beyond the understanding of QND with simpler EPs~\cite{wang2019non}; we will address this aspect in a future publication.
Furthermore, the cusps of the swallowtail mark the transition from exceptional to diabolical degeneracies when linear coupling reduces to zero at the three-fold diabolical line (DL3). 

\begin{figure}
	\centering
	\includegraphics[width=1\linewidth]{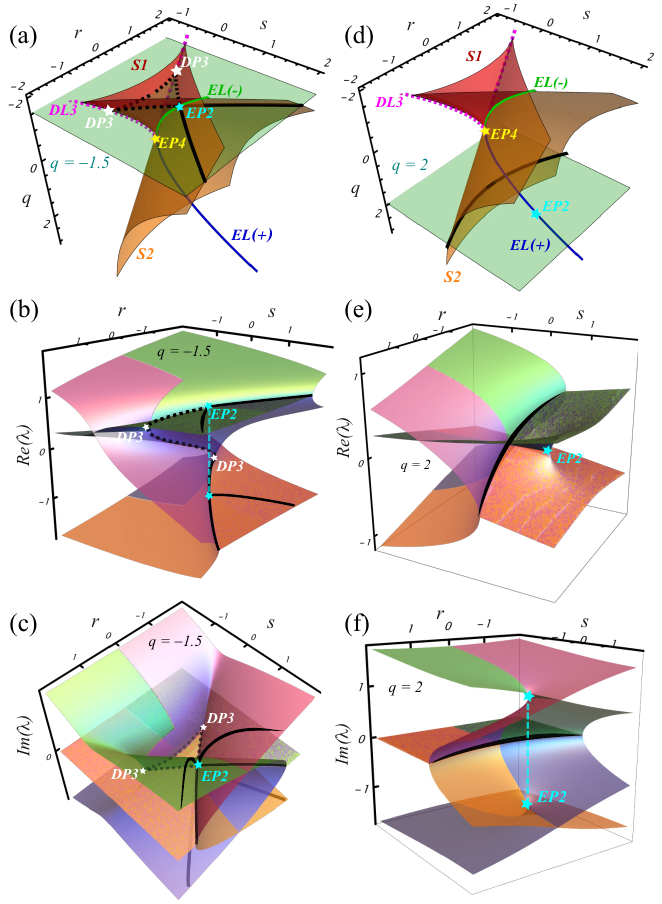}
	\caption{Eigenvalue dependence on representative $\mathbb{R}^2$ subspace of the control parameter space. (a-c) $q=-1.5$ plane intersecting the swallowtail, followed by plots of (b) real and (c) imaginary parts of the four eigenvalues $\lambda$. (d-f) Analogous plots for $q=2$ plane. 
	}
	\label{fig::planes}
\end{figure}
\begin{figure*}
	\centering
	\includegraphics[width=1\textwidth]{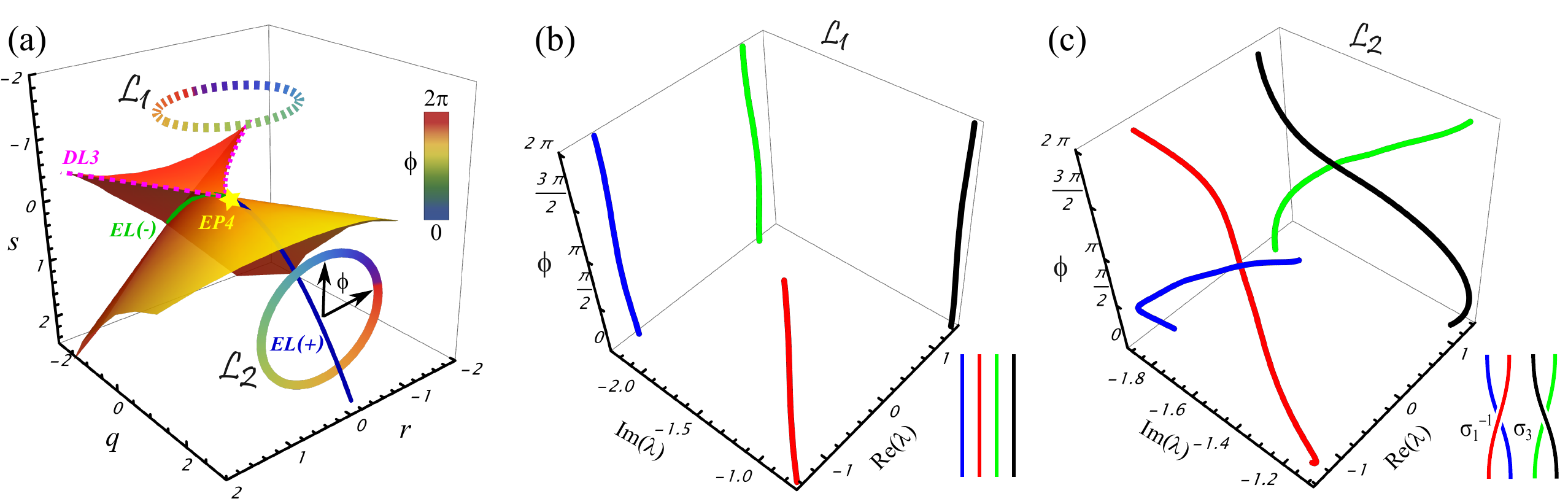}
	\caption{Forming nontrivial braids with control parameter loops. (a) 
		Two control loops in the control parameter space that do not go through any degeneracy: $\mathcal{L}_1$ (dashed) not enclosing any degeneracy, and $\mathcal{L}_2$ (solid) enclosing the EL($+$). The trajectories are parametrized by angle $\phi$, which ranges from $0$ to $2\pi$. (b-c) Plots of real and imaginary parts of the eigenvalues $\lambda$ and their artin braid representation along loops (b) $\mathcal{L}_1$ and (c) $\mathcal{L}_2$.
	}
	\label{braids}
\end{figure*}
We highlight some of the important features of the non-Hermitian degeneracies on the swallowtail.
To illustrate these features, we focus on the $q=-1.5$ [Fig. \hyperref[fig::planes]{2(a)}] and $q=2$ [Fig. \hyperref[fig::planes]{2(d)}] planes in the control parameter space that intersect the swallowtail and plot the corresponding real [Figs. \hyperref[fig::planes]{2(b,e)}] and imaginary [Figs. \hyperref[fig::planes]{2(c,f)}] parts of the four eigenvalues $\lambda$ (see also Supplemental Videos 1-2~\cite{supp}). The $q=-1.5$ plane intersects both S1 (dotted black lines) and S2 (solid black lines) in the parameter space [Fig. \hyperref[fig::planes]{2(a)}], and we plot the corresponding trajectories on the eigenvalue plots [Figs. \hyperref[fig::planes]{2(b,c)}] when traversing these lines. 
The pairwise second-order EP (EP2) marked by a cyan star in Fig. \hyperref[fig::planes]{2(a)} is an important transition point. On the positive $r$ part of surface S1, along the line from the three-fold diabolical degeneracy (DP3, marked by white star) to EP2,
there is one doubly-degenerate real eigenvalue (blue/green) and two distinct real eigenvalues (pink, orange). The degenerate eigenvector $|\mathbf{R}\rangle$ satisfies $|\mathbf{R}\rangle \propto \tau_x|\mathbf{R}\rangle^\ast.$
From EP2 along the line on surface S2 for positive $r$, the formerly degenerate real eigenvalues split into a complex conjugate pair, and hence such a linear dependence is broken, i.e., the degenerate eigenvector splits into two linearly independent vectors $|\mathbf{R}\rangle$ and $\tau_x|\mathbf{R}\rangle^\ast$. 
This phenomenon is known as spontaneous symmetry breaking~\cite{melkani}, and specifically here, it is a spontaneous breaking of the particle-hole symmetry. 
Conversely, the formerly distinct real eigenvalues (orange, pink)  merge into a purely real degenerate pair after EP2; their eigenvectors coalesce and satisfy $|\mathbf{R}\rangle \propto \tau_x|\mathbf{R}\rangle^\ast.$ While this discussion is based on the $q=-1.5$ plane, it can be seen that on the swallowtail, the entire EL($-$) marks such a spontaneous symmetry breaking.  

Going from the $q=-1.5$ plane towards the $q=0$ plane, one can see that the triangular intersection formed by two DP3's and the EP2 shrinks, and this triangle collapses into one EP4. At positive $q$ values, e.g., on the $q=2$ plane that we show in Fig. \hyperref[fig::planes]{2(d)}, the outgoing EL($+$) intersects with the plane as a separate EP, denoted as EP2 (cyan star). At the EP2, there are two doubly-degenerate eigenvalues that are complex conjugate with each other. 
On the other hand, the intersecting line of S2 with the $q=2$ plane still hosts a two-fold real degeneracy and two other distinct eigenvalues that form a complex conjugate pair [Figs. \hyperref[fig::planes]{(e,f)}]. For different views of Figs. \hyperref[fig::planes]{(e,f)} see also Supplemental Videos 3-4~\cite{supp}.

\textit{Eigenvalue braids.}---A striking manifestation of the swallowtail degeneracy structure in bosonic quadratic systems is the possibility of forming nontrivial braids with complex eigenvalues for a simple control parameter loop. 
It has been shown that in classical non-Hermitian systems, when control parameters are varied along a loop that encloses a non-Hermitian degeneracy without crossing any degeneracies, the complex eigenspectrum may return to itself in a topologically nontrivial way \cite{wang2021topological,patil2022measuring} by forming a braid. However, from a simple codimension argument, it may seem impossible for the eigenvalues of two-mode squeezing dynamical matrices to achieve such braids. Indeed, if one looks at the PHP associated with a two-mode closed squeezing system, the codimension of the degeneracy structure in the control parameter space is one as the degeneracies form lines in $\mathbb{R}^2$. By contrast, a nontrivial braid can be formed with a codimension two, e.g., an isolated EP in $\mathbb{R}^2$ or an isolated or partially isolated exceptional line in $\mathbb{R}^3$. Here, by breaking the pseudo-Hermiticity through judiciously added losses in such a two-mode squeezing system, the rich swallowtail degeneracy structure that we identify is an enabling factor for forming nontrivial braids. More specifically, the outgoing section of the exceptional line EL($+$) has a codimension two in the $\mathbb{R}^3$ control parameter space, enabling us to create control loops that form nontrivial braids. 

As an example, in Fig. \hyperref[braids]{3(a)} we draw two control parameter loops in such a $\mathbb{R}^3$ parameter space (specific parameters associated with the loops are given in Supplemental Sec. II B~\cite{supp}). 
One loop, namely $\mathcal{L}_1$, denoted by the dashed circle in Fig.~\hyperref[braids]{3(a)} does not enclose any degeneracy. In Fig. \hyperref[braids]{3(b)}, we plot the four complex eigenvalues as a function of the parametrization angle $\phi$ associated with the loop in the coordinate of [Re($\lambda$), Im($\lambda$), $\phi$] space. One can see that by traversing the loop $\mathcal{L}_1$, the four eigenvalues trivially go back to the initial values with no permutation. Mathematically, such a behavior can be denoted as a trivial braid diagram [bottom right of Fig. \hyperref[braids]{3(b)}]. By contrast, the other control parameter loop, namely $\mathcal{L}_2$, encloses the exceptional line EL($+$) that goes out of the swallowtail's surface. The two loops are not homotopy equivalent, i.e., $\mathcal{L}_2$ cannot be continuously transformed into $\mathcal{L}_1$ without crossing a degeneracy. As we show an analogous plot for the eigenvalues along $\mathcal{L}_2$ in Fig. \hyperref[braids]{3(c)} in the [Re($\lambda$), Im($\lambda$), $\phi$] space, one can see that there is a half twist for each pair of eigenvalues, leading to a nontrivial permutation at the end. Red and blue, as well as green and black eigenvalue "strings" switch in pairs, as dictated by the particle-hole symmetry. Such a structure can be considered as a nontrivial braid in the $\mathbb{B}_4$ group, shown by the braid diagram in the bottom right of Fig. \hyperref[braids]{3(c)}] corresponding to the braid word $\sigma_1^{-1} \sigma_3$. This example showcases the remarkable qualitative difference between a driven-dissipative squeezing and a closed-system squeezing and the important implication of identifying the swallowtail degeneracy structure.

In summary, we identified the swallowtail structure consisting of non-Hermitian and Hermitian degeneracies in a two-mode quadrature squeezing system's dynamical matrix. We find that the enabling factor that gives rise to the swallowtail degeneracy is the breaking of the pseudo-Hermiticity of the dynamical matrix, which can be achieved by having asymmetric losses on the two modes. The interplay among driving (squeezing), dissipation (losses), and linear coupling (beam-splitter-type interaction) collectively give rise to the rich degeneracy morphology of the swallowtail. We also discover that such a rich degeneracy structure enables the formation of nontrivial eigenvalue braids by looping around a doubly degenerate exceptional line, which is not possible in two-mode squeezing systems without nontrivial losses. Our model is readily implementable in optical or photonic experiments, as the simple model we provide only needs single-mode squeezing accompanied by linear coupling. Looking ahead, one may ask if a dynamic encirclement of the EL can lead to exotic effects, such as topological chiral gates of Gaussian states. The characterization of this rich degeneracy morphology provides important insights into the dynamics of quantum states of light in quadratic systems, and we anticipate future extensions to multimode squeezing systems can lead to new paradigms of Gaussian state generation and control.   

\begin{acknowledgments}

We acknowledge the support from Qu\'ebec’s Minist\`ere de l’\'Economie, de l’Innovation et de l’ \'Energie (MEIE), Photonique Quantique Qu\'ebec (PQ2), and Natural Sciences and Engineering Research Council of Canada (NSERC), [RGPIN-2023-03630].
\end{acknowledgments}

\bibliographystyle{aipAuAll_links}

\begin{thebibliography}{39}%
	\makeatletter
	\providecommand \@ifxundefined [1]{%
		\@ifx{#1\undefined}
	}%
	\providecommand \@ifnum [1]{%
		\ifnum #1\expandafter \@firstoftwo
		\else \expandafter \@secondoftwo
		\fi
	}%
	\providecommand \@ifx [1]{%
		\ifx #1\expandafter \@firstoftwo
		\else \expandafter \@secondoftwo
		\fi
	}%
	\providecommand \natexlab [1]{#1}%
	\providecommand \enquote  [1]{``#1''}%
	\providecommand \bibnamefont  [1]{#1}%
	\providecommand \bibfnamefont [1]{#1}%
	\providecommand \citenamefont [1]{#1}%
	\providecommand \href@noop [0]{\@secondoftwo}%
	\providecommand \href [0]{\begingroup \@sanitize@url \@href}%
	\providecommand \@href[1]{\@@startlink{#1}\@@href}%
	\providecommand \@@href[1]{\endgroup#1\@@endlink}%
	\providecommand \@sanitize@url [0]{\catcode `\\12\catcode `\$12\catcode
		`\&12\catcode `\#12\catcode `\^12\catcode `\_12\catcode `\%12\relax}%
	\providecommand \@@startlink[1]{}%
	\providecommand \@@endlink[0]{}%
	\providecommand \url  [0]{\begingroup\@sanitize@url \@url }%
	\providecommand \@url [1]{\endgroup\@href {#1}{\urlprefix }}%
	\providecommand \urlprefix  [0]{URL }%
	\providecommand \Eprint [0]{\href }%
	\providecommand \doibase [0]{http://dx.doi.org/}%
	\providecommand \selectlanguage [0]{\@gobble}%
	\providecommand \bibinfo  [0]{\@secondoftwo}%
	\providecommand \bibfield  [0]{\@secondoftwo}%
	\providecommand \translation [1]{[#1]}%
	\providecommand \BibitemOpen [0]{}%
	\providecommand \bibitemStop [0]{}%
	\providecommand \bibitemNoStop [0]{.\EOS\space}%
	\providecommand \EOS [0]{\spacefactor3000\relax}%
	\providecommand \BibitemShut  [1]{\csname bibitem#1\endcsname}%
	\let\auto@bib@innerbib\@empty
	\bibitem {arnol2003catastrophe}%
	\BibitemOpen
	\bibfield  {author} {\bibinfo {author} {\bibfnamefont {V.~I.}\ \bibnamefont
			{Arnold}},\ }\href@noop {} {\emph {\bibinfo {title} {Catastrophe theory}}}\
	(\bibinfo  {publisher} {Springer Science \& Business Media},\ \bibinfo {year}
	{2003})\BibitemShut {NoStop}%
	\bibitem {arnoldMHD83}%
	\BibitemOpen
	\bibfield  {author} {\bibinfo {author} {\bibfnamefont {V.~I.}\ \bibnamefont
			{{Arnold}}}\ and\ \bibinfo {author} {\bibfnamefont {E.~I.}\ \bibnamefont
			{{Korkina}}},\ }\href@noop {} {\bibfield  {journal} {\bibinfo  {journal}
			{Vestnik Moskov. Univ. Ser.~1. Mat. Mekh.}\ }\textbf {\bibinfo {volume}
			{3}},\ \bibinfo {pages} {43} (\bibinfo {year} {1983})}\BibitemShut {NoStop}%
	\bibitem {Raz2012}%
	\BibitemOpen
	\bibfield  {author} {\bibinfo {author} {\bibfnamefont {O.}~\bibnamefont
			{Raz}}, \bibinfo {author} {\bibfnamefont {O.}~\bibnamefont {Pedatzur}},
		\bibinfo {author} {\bibfnamefont {B.~D.}\ \bibnamefont {Bruner}}, \ and\
		\bibinfo {author} {\bibfnamefont {N.}~\bibnamefont {Dudovich}},\ }\href
	{\doibase 10.1038/nphoton.2011.353} {\bibfield  {journal} {\bibinfo
			{journal} {Nature Photonics}\ }\textbf {\bibinfo {volume} {6}},\ \bibinfo
		{pages} {170} (\bibinfo {year} {2012})}\BibitemShut {NoStop}%
	\bibitem {ding2022non}%
	\BibitemOpen
	\bibfield  {author} {\bibinfo {author} {\bibfnamefont {K.}~\bibnamefont
			{Ding}}, \bibinfo {author} {\bibfnamefont {C.}~\bibnamefont {Fang}}, \ and\
		\bibinfo {author} {\bibfnamefont {G.}~\bibnamefont {Ma}},\ }\href {\doibase
		https://doi.org/10.1038/s42254-022-00516-5} {\bibfield  {journal} {\bibinfo
			{journal} {Nature Reviews Physics}\ }\textbf {\bibinfo {volume} {4}},\
		\bibinfo {pages} {745} (\bibinfo {year} {2022})}\BibitemShut {NoStop}%
	\bibitem {heiss2012physics}%
	\BibitemOpen
	\bibfield  {author} {\bibinfo {author} {\bibfnamefont {W.~D.}\ \bibnamefont
			{Heiss}},\ }\href {\doibase 10.1088/1751-8113/45/44/444016} {\bibfield
		{journal} {\bibinfo  {journal} {J. Phys. A: Math. Theor.}\ }\textbf {\bibinfo
			{volume} {45}},\ \bibinfo {pages} {444016} (\bibinfo {year}
		{2012})}\BibitemShut {NoStop}%
	\bibitem {ozdemir2019parity}%
	\BibitemOpen
	\bibfield  {author} {\bibinfo {author} {\bibfnamefont {{\c{S}}.~K.}\
			\bibnamefont {{\"O}zdemir}}, \bibinfo {author} {\bibfnamefont
			{S.}~\bibnamefont {Rotter}}, \bibinfo {author} {\bibfnamefont
			{F.}~\bibnamefont {Nori}}, \ and\ \bibinfo {author} {\bibfnamefont
			{L.}~\bibnamefont {Yang}},\ }\href {\doibase 10.1038/s41563-019-0304-9}
	{\bibfield  {journal} {\bibinfo  {journal} {Nature materials}\ }\textbf
		{\bibinfo {volume} {18}},\ \bibinfo {pages} {783} (\bibinfo {year}
		{2019})}\BibitemShut {NoStop}%
	\bibitem {parto2020non}%
	\BibitemOpen
	\bibfield  {author} {\bibinfo {author} {\bibfnamefont {M.}~\bibnamefont
			{Parto}}, \bibinfo {author} {\bibfnamefont {Y.~G.}\ \bibnamefont {Liu}},
		\bibinfo {author} {\bibfnamefont {B.}~\bibnamefont {Bahari}}, \bibinfo
		{author} {\bibfnamefont {M.}~\bibnamefont {Khajavikhan}}, \ and\ \bibinfo
		{author} {\bibfnamefont {D.~N.}\ \bibnamefont {Christodoulides}},\ }\href
	{\doibase 10.1515/nanoph-2020-0434} {\bibfield  {journal} {\bibinfo
			{journal} {Nanophotonics}\ }\textbf {\bibinfo {volume} {10}},\ \bibinfo
		{pages} {403} (\bibinfo {year} {2020})}\BibitemShut {NoStop}%
	\bibitem {Hu2023}%
	\BibitemOpen
	\bibfield  {author} {\bibinfo {author} {\bibfnamefont {J.}~\bibnamefont
			{Hu}}, \bibinfo {author} {\bibfnamefont {R.-Y.}\ \bibnamefont {Zhang}},
		\bibinfo {author} {\bibfnamefont {Y.}~\bibnamefont {Wang}}, \bibinfo {author}
		{\bibfnamefont {X.}~\bibnamefont {Ouyang}}, \bibinfo {author} {\bibfnamefont
			{Y.}~\bibnamefont {Zhu}}, \bibinfo {author} {\bibfnamefont {H.}~\bibnamefont
			{Jia}}, \ and\ \bibinfo {author} {\bibfnamefont {C.~T.}\ \bibnamefont
			{Chan}},\ }\href {\doibase 10.1038/s41567-023-02048-w} {\bibfield  {journal}
		{\bibinfo  {journal} {Nature Physics}\ }\textbf {\bibinfo {volume} {19}},\
		\bibinfo {pages} {1098} (\bibinfo {year} {2023})}\BibitemShut {NoStop}%
	\bibitem {KatsuraPhysRevLett.104.066403}%
	\BibitemOpen
	\bibfield  {author} {\bibinfo {author} {\bibfnamefont {H.}~\bibnamefont
			{Katsura}}, \bibinfo {author} {\bibfnamefont {N.}~\bibnamefont {Nagaosa}}, \
		and\ \bibinfo {author} {\bibfnamefont {P.~A.}\ \bibnamefont {Lee}},\ }\href
	{\doibase 10.1103/PhysRevLett.104.066403} {\bibfield  {journal} {\bibinfo
			{journal} {Phys. Rev. Lett.}\ }\textbf {\bibinfo {volume} {104}},\ \bibinfo
		{pages} {066403} (\bibinfo {year} {2010})}\BibitemShut {NoStop}%
	\bibitem {tserkovnyak}%
	\BibitemOpen
	\bibfield  {author} {\bibinfo {author} {\bibfnamefont {S.~K.}\ \bibnamefont
			{Kim}}, \bibinfo {author} {\bibfnamefont {H.}~\bibnamefont {Ochoa}}, \bibinfo
		{author} {\bibfnamefont {R.}~\bibnamefont {Zarzuela}}, \ and\ \bibinfo
		{author} {\bibfnamefont {Y.}~\bibnamefont {Tserkovnyak}},\ }\href {\doibase
		10.1103/PhysRevLett.117.227201} {\bibfield  {journal} {\bibinfo  {journal}
			{Phys. Rev. Lett.}\ }\textbf {\bibinfo {volume} {117}},\ \bibinfo {pages}
		{227201} (\bibinfo {year} {2016})}\BibitemShut {NoStop}%
	\bibitem {owerre2016first}%
	\BibitemOpen
	\bibfield  {author} {\bibinfo {author} {\bibfnamefont {S.}~\bibnamefont
			{Owerre}},\ }\href {\doibase 10.1088/0953-8984/28/38/386001} {\bibfield
		{journal} {\bibinfo  {journal} {Journal of Physics: Condensed Matter}\
		}\textbf {\bibinfo {volume} {28}},\ \bibinfo {pages} {386001} (\bibinfo
		{year} {2016})}\BibitemShut {NoStop}%
	\bibitem {walls1983squeezed}%
	\BibitemOpen
	\bibfield  {author} {\bibinfo {author} {\bibfnamefont {D.~F.}\ \bibnamefont
			{Walls}},\ }\href {\doibase 10.1038/306141a0} {\bibfield  {journal} {\bibinfo
			{journal} {Nature}\ }\textbf {\bibinfo {volume} {306}},\ \bibinfo {pages}
		{141} (\bibinfo {year} {1983})}\BibitemShut {NoStop}%
	\bibitem {schumakerPhysRevA.31.3068}%
	\BibitemOpen
	\bibfield  {author} {\bibinfo {author} {\bibfnamefont {C.~M.}\ \bibnamefont
			{Caves}}\ and\ \bibinfo {author} {\bibfnamefont {B.~L.}\ \bibnamefont
			{Schumaker}},\ }\href {\doibase 10.1103/PhysRevA.31.3068} {\bibfield
		{journal} {\bibinfo  {journal} {Phys. Rev. A}\ }\textbf {\bibinfo {volume}
			{31}},\ \bibinfo {pages} {3068} (\bibinfo {year} {1985})}\BibitemShut
	{NoStop}%
	\bibitem {andersen201630}%
	\BibitemOpen
	\bibfield  {author} {\bibinfo {author} {\bibfnamefont {U.~L.}\ \bibnamefont
			{Andersen}}, \bibinfo {author} {\bibfnamefont {T.}~\bibnamefont {Gehring}},
		\bibinfo {author} {\bibfnamefont {C.}~\bibnamefont {Marquardt}}, \ and\
		\bibinfo {author} {\bibfnamefont {G.}~\bibnamefont {Leuchs}},\ }\href
	{\doibase 10.1088/0031-8949/91/5/053001} {\bibfield  {journal} {\bibinfo
			{journal} {Physica Scripta}\ }\textbf {\bibinfo {volume} {91}},\ \bibinfo
		{pages} {053001} (\bibinfo {year} {2016})}\BibitemShut {NoStop}%
	\bibitem {caves}%
	\BibitemOpen
	\bibfield  {author} {\bibinfo {author} {\bibfnamefont {C.~M.}\ \bibnamefont
			{Caves}},\ }\href {\doibase 10.1103/PhysRevD.23.1693} {\bibfield  {journal}
		{\bibinfo  {journal} {Phys. Rev. D}\ }\textbf {\bibinfo {volume} {23}},\
		\bibinfo {pages} {1693} (\bibinfo {year} {1981})}\BibitemShut {NoStop}%
	\bibitem {aasi2013enhanced}%
	\BibitemOpen
	\bibfield  {author} {\bibinfo {author} {\bibfnamefont {J.}~\bibnamefont
			{Aasi}}, \bibinfo {author} {\bibfnamefont {J.}~\bibnamefont {Abadie}},
		\bibinfo {author} {\bibfnamefont {B.}~\bibnamefont {Abbott}}, \bibinfo
		{author} {\bibfnamefont {R.}~\bibnamefont {Abbott}}, \bibinfo {author}
		{\bibfnamefont {T.}~\bibnamefont {Abbott}}, \bibinfo {author} {\bibfnamefont
			{M.}~\bibnamefont {Abernathy}}, \bibinfo {author} {\bibfnamefont
			{C.}~\bibnamefont {Adams}}, \bibinfo {author} {\bibfnamefont
			{T.}~\bibnamefont {Adams}}, \bibinfo {author} {\bibfnamefont
			{P.}~\bibnamefont {Addesso}}, \bibinfo {author} {\bibfnamefont
			{R.}~\bibnamefont {Adhikari}},  \emph {et~al.},\ }\href {\doibase
		https://doi.org/10.1038/nphoton.2013.177} {\bibfield  {journal} {\bibinfo
			{journal} {Nature Photonics}\ }\textbf {\bibinfo {volume} {7}},\ \bibinfo
		{pages} {613} (\bibinfo {year} {2013})}\BibitemShut {NoStop}%
	\bibitem {braunsteinRevModPhys.77.513}%
	\BibitemOpen
	\bibfield  {author} {\bibinfo {author} {\bibfnamefont {S.~L.}\ \bibnamefont
			{Braunstein}}\ and\ \bibinfo {author} {\bibfnamefont {P.}~\bibnamefont {van
				Loock}},\ }\href {\doibase 10.1103/RevModPhys.77.513} {\bibfield  {journal}
		{\bibinfo  {journal} {Rev. Mod. Phys.}\ }\textbf {\bibinfo {volume} {77}},\
		\bibinfo {pages} {513} (\bibinfo {year} {2005})}\BibitemShut {NoStop}%
	\bibitem {shapiro1055958}%
	\BibitemOpen
	\bibfield  {author} {\bibinfo {author} {\bibfnamefont {H.}~\bibnamefont
			{Yuen}}\ and\ \bibinfo {author} {\bibfnamefont {J.}~\bibnamefont {Shapiro}},\
	}\href {\doibase 10.1109/TIT.1978.1055958} {\bibfield  {journal} {\bibinfo
			{journal} {IEEE Transactions on Information Theory}\ }\textbf {\bibinfo
			{volume} {24}},\ \bibinfo {pages} {657} (\bibinfo {year} {1978})}\BibitemShut
	{NoStop}%
	\bibitem {wang2019non}%
	\BibitemOpen
	\bibfield  {author} {\bibinfo {author} {\bibfnamefont {Y.-X.}\ \bibnamefont
			{Wang}}\ and\ \bibinfo {author} {\bibfnamefont {A.~A.}~\bibnamefont {Clerk}},\
	}\href {\doibase https://doi.org/10.1103/PhysRevA.99.063834} {\bibfield
		{journal} {\bibinfo  {journal} {Physical Review A}\ }\textbf {\bibinfo
			{volume} {99}},\ \bibinfo {pages} {063834} (\bibinfo {year}
		{2019})}\BibitemShut {NoStop}%
	\bibitem {PhysRevX.8.041031}%
	\BibitemOpen
	\bibfield  {author} {\bibinfo {author} {\bibfnamefont {A.}~\bibnamefont
			{McDonald}}, \bibinfo {author} {\bibfnamefont {T.}~\bibnamefont
			{Pereg-Barnea}}, \ and\ \bibinfo {author} {\bibfnamefont {A.~A.}\
			\bibnamefont {Clerk}},\ }\href {\doibase 10.1103/PhysRevX.8.041031}
	{\bibfield  {journal} {\bibinfo  {journal} {Phys. Rev. X}\ }\textbf {\bibinfo
			{volume} {8}},\ \bibinfo {pages} {041031} (\bibinfo {year}
		{2018})}\BibitemShut {NoStop}%
	\bibitem {flynn2020deconstructing}%
	\BibitemOpen
	\bibfield  {author} {\bibinfo {author} {\bibfnamefont {V.~P.}\ \bibnamefont
			{Flynn}}, \bibinfo {author} {\bibfnamefont {E.}~\bibnamefont {Cobanera}}, \
		and\ \bibinfo {author} {\bibfnamefont {L.}~\bibnamefont {Viola}},\ }\href
	{\doibase 10.1088/1367-2630/ab9e87} {\bibfield  {journal} {\bibinfo
			{journal} {New Journal of Physics}\ }\textbf {\bibinfo {volume} {22}},\
		\bibinfo {pages} {083004} (\bibinfo {year} {2020})}\BibitemShut {NoStop}%
	\bibitem {Roy:21}%
	\BibitemOpen
	\bibfield  {author} {\bibinfo {author} {\bibfnamefont {A.}~\bibnamefont
			{Roy}}, \bibinfo {author} {\bibfnamefont {S.}~\bibnamefont {Jahani}},
		\bibinfo {author} {\bibfnamefont {Q.}~\bibnamefont {Guo}}, \bibinfo {author}
		{\bibfnamefont {A.}~\bibnamefont {Dutt}}, \bibinfo {author} {\bibfnamefont
			{S.}~\bibnamefont {Fan}}, \bibinfo {author} {\bibfnamefont {M.-A.}\
			\bibnamefont {Miri}}, \ and\ \bibinfo {author} {\bibfnamefont
			{A.}~\bibnamefont {Marandi}},\ }\href {\doibase 10.1364/OPTICA.415569}
	{\bibfield  {journal} {\bibinfo  {journal} {Optica}\ }\textbf {\bibinfo
			{volume} {8}},\ \bibinfo {pages} {415} (\bibinfo {year} {2021})}\BibitemShut
	{NoStop}%
	\bibitem {ShengwangPhysRevLett.128.173602}%
	\BibitemOpen
	\bibfield  {author} {\bibinfo {author} {\bibfnamefont {X.-W.}\ \bibnamefont
			{Luo}}, \bibinfo {author} {\bibfnamefont {C.}~\bibnamefont {Zhang}}, \ and\
		\bibinfo {author} {\bibfnamefont {S.}~\bibnamefont {Du}},\ }\href {\doibase
		10.1103/PhysRevLett.128.173602} {\bibfield  {journal} {\bibinfo  {journal}
			{Phys. Rev. Lett.}\ }\textbf {\bibinfo {volume} {128}},\ \bibinfo {pages}
		{173602} (\bibinfo {year} {2022})}\BibitemShut {NoStop}%
	\bibitem {Braunstein}%
	\BibitemOpen
	\bibfield  {author} {\bibinfo {author} {\bibfnamefont {S.~L.}\ \bibnamefont
			{Braunstein}},\ }\href {\doibase 10.1103/PhysRevA.71.055801} {\bibfield
		{journal} {\bibinfo  {journal} {Phys. Rev. A}\ }\textbf {\bibinfo {volume}
			{71}},\ \bibinfo {pages} {055801} (\bibinfo {year} {2005})}\BibitemShut
	{NoStop}%
	\bibitem {adesso2014continuous}%
	\BibitemOpen
	\bibfield  {author} {\bibinfo {author} {\bibfnamefont {G.}~\bibnamefont
			{Adesso}}, \bibinfo {author} {\bibfnamefont {S.}~\bibnamefont {Ragy}}, \ and\
		\bibinfo {author} {\bibfnamefont {A.~R.}\ \bibnamefont {Lee}},\ }\href
	{\doibase 10.1142/S1230161214400010} {\bibfield  {journal} {\bibinfo
			{journal} {Open Systems \& Information Dynamics}\ }\textbf {\bibinfo {volume}
			{21}},\ \bibinfo {pages} {1440001} (\bibinfo {year} {2014})}\BibitemShut
	{NoStop}%
	\bibitem {supp}%
	\BibitemOpen
	\href@noop {} {}\bibinfo {note} {See Supplemental Material at URL for details
		and videos, which includes Refs. [27-33].}\BibitemShut {Stop}%
	\bibitem {Walls2008}%
	\BibitemOpen
	\bibfield  {author} {\bibinfo {author} {\bibfnamefont {D.}~\bibnamefont
			{Walls}}\ and\ \bibinfo {author} {\bibfnamefont {G.~J.}\ \bibnamefont
			{Milburn}},\ }\enquote {\bibinfo {title} {Input--output formulation of
			optical cavities},}\ in\ \href@noop {} {\emph {\bibinfo {booktitle} {Quantum
				Optics}}},\ \bibinfo {editor} {edited by\ \bibinfo {editor} {\bibfnamefont
			{D.}~\bibnamefont {Walls}}\ and\ \bibinfo {editor} {\bibfnamefont {G.~J.}\
			\bibnamefont {Milburn}}}\ (\bibinfo  {publisher} {Springer Berlin
		Heidelberg},\ \bibinfo {address} {Berlin, Heidelberg},\ \bibinfo {year}
	{2008})\ pp.\ \bibinfo {pages} {127--141}\BibitemShut {NoStop}%
	\bibitem {michael_spivak_comprehensive_1979}%
	\BibitemOpen
	\bibfield  {author} {\bibinfo {author} {\bibfnamefont {M.}~\bibnamefont
			{Spivak}},\ }\href@noop {} {\emph {\bibinfo {title} {A Comprehensive
				Introduction to Differential Geometry}}},\ \bibinfo {edition} {3rd}\ ed.,\
	Vol.~\bibinfo {volume} {1}\ (\bibinfo  {publisher} {Publish Or Perish},\
	\bibinfo {year} {1999})\BibitemShut {NoStop}%
	\bibitem {melkani}%
	\BibitemOpen
	\bibfield  {author} {\bibinfo {author} {\bibfnamefont {A.}~\bibnamefont
			{Melkani}},\ }\href {\doibase 10.1103/PhysRevResearch.5.023035} {\bibfield
		{journal} {\bibinfo  {journal} {Phys. Rev. Res.}\ }\textbf {\bibinfo {volume}
			{5}},\ \bibinfo {pages} {023035} (\bibinfo {year} {2023})}\BibitemShut
	{NoStop}%
	\bibitem {vassiliev1995topology}%
	\BibitemOpen
	\bibfield  {author} {\bibinfo {author} {\bibfnamefont {V.~A.}\ \bibnamefont
			{Vassiliev}},\ }in\ \href@noop {} {\emph {\bibinfo {booktitle} {Proceedings
				of the International Congress of Mathematicians: August 3--11, 1994
				Z{\"u}rich, Switzerland}}}\ (\bibinfo {organization} {Springer},\ \bibinfo
	{year} {1995})\ pp.\ \bibinfo {pages} {209--226}\BibitemShut {NoStop}%
	\bibitem {batorova2013desingularization}%
	\BibitemOpen
	\bibfield  {author} {\bibinfo {author} {\bibfnamefont {M.}~\bibnamefont
			{B{\'a}torov{\'a}}}, \bibinfo {author} {\bibfnamefont {M.}~\bibnamefont
			{Val{\'\i}kov{\'a}}}, \ and\ \bibinfo {author} {\bibfnamefont
			{P.}~\bibnamefont {Chalmoviansk{\`y}}},\ }in\ \href@noop {} {\emph {\bibinfo
			{booktitle} {Proceedings of the 29th Spring Conference on Computer
				Graphics}}}\ (\bibinfo {year} {2013})\ pp.\ \bibinfo {pages}
	{35--42}\BibitemShut {NoStop}%
	\bibitem {spivak2018calculus}%
	\BibitemOpen
	\bibfield  {author} {\bibinfo {author} {\bibfnamefont {M.}~\bibnamefont
			{Spivak}},\ }\href@noop {} {\emph {\bibinfo {title} {Calculus on manifolds: a
				modern approach to classical theorems of advanced calculus}}}\ (\bibinfo
	{publisher} {CRC press},\ \bibinfo {year} {2018})\BibitemShut {NoStop}%
	\bibitem {ARNON198837}%
	\BibitemOpen
	\bibfield  {author} {\bibinfo {author} {\bibfnamefont {D.~S.}\ \bibnamefont
			{Arnon}},\ }\href {\doibase 10.1016/0004-3702(88)90049-5} {\bibfield
		{journal} {\bibinfo  {journal} {Artificial Intelligence}\ }\textbf {\bibinfo
			{volume} {37}},\ \bibinfo {pages} {37} (\bibinfo {year} {1988})}\BibitemShut
	{NoStop}%
	\bibitem {Gardiner1985}%
	\BibitemOpen
	\bibfield  {author} {\bibinfo {author} {\bibfnamefont {C.~W.}\ \bibnamefont
			{Gardiner}}\ and\ \bibinfo {author} {\bibfnamefont {M.~J.}\ \bibnamefont
			{Collett}},\ }\href {\doibase 10.1103/PhysRevA.31.3761} {\bibfield  {journal}
		{\bibinfo  {journal} {Phys. Rev. A}\ }\textbf {\bibinfo {volume} {31}},\
		\bibinfo {pages} {3761} (\bibinfo {year} {1985})}\BibitemShut {NoStop}%
	\bibitem {phs}%
	\BibitemOpen
	\bibfield  {author} {\bibinfo {author} {\bibfnamefont {H.}~\bibnamefont
			{Zhou}}\ and\ \bibinfo {author} {\bibfnamefont {J.~Y.}\ \bibnamefont {Lee}},\
	}\href {\doibase 10.1103/PhysRevB.99.235112} {\bibfield  {journal} {\bibinfo
			{journal} {Phys. Rev. B}\ }\textbf {\bibinfo {volume} {99}},\ \bibinfo
		{pages} {235112} (\bibinfo {year} {2019})}\BibitemShut {NoStop}%
	\bibitem {Mostafazadeh}%
	\BibitemOpen
	\bibfield  {author} {\bibinfo {author} {\bibfnamefont {A.}~\bibnamefont
			{Mostafazadeh}},\ }\href {\doibase 10.1063/1.1418246} {\bibfield  {journal}
		{\bibinfo  {journal} {Journal of Mathematical Physics}\ }\textbf {\bibinfo
			{volume} {43}},\ \bibinfo {pages} {205} (\bibinfo {year} {2002})}\BibitemShut
	{NoStop}%
	\bibitem {arnold2006mathematical}%
	\BibitemOpen
	\bibfield  {author} {\bibinfo {author} {\bibfnamefont {V.~I.}\ \bibnamefont
			{Arnold}}, \bibinfo {author} {\bibfnamefont {A.~A.}\ \bibnamefont {Davydov}},
		\bibinfo {author} {\bibfnamefont {V.~A.}\ \bibnamefont {Vassiliev}}, \ and\
		\bibinfo {author} {\bibfnamefont {V.}~\bibnamefont {Zakalyukin}},\
	}\href@noop {} {\emph {\bibinfo {title} {Mathematical Models of Catastrophes:
				Control of Catastrophic Processes}}},\ IIASA reprint\ (\bibinfo  {publisher}
	{Internat. Inst. for Applied Systems Analysis},\ \bibinfo {year}
	{2006})\BibitemShut {NoStop}%
	\bibitem {wang2021topological}%
	\BibitemOpen
	\bibfield  {author} {\bibinfo {author} {\bibfnamefont {K.}~\bibnamefont
			{Wang}}, \bibinfo {author} {\bibfnamefont {A.}~\bibnamefont {Dutt}}, \bibinfo
		{author} {\bibfnamefont {C.~C.}\ \bibnamefont {Wojcik}}, \ and\ \bibinfo
		{author} {\bibfnamefont {S.}~\bibnamefont {Fan}},\ }\href {\doibase
		10.1038/s41586-021-03848-x} {\bibfield  {journal} {\bibinfo  {journal}
			{Nature}\ }\textbf {\bibinfo {volume} {598}},\ \bibinfo {pages} {59}
		(\bibinfo {year} {2021})}\BibitemShut {NoStop}%
	\bibitem {patil2022measuring}%
	\BibitemOpen
	\bibfield  {author} {\bibinfo {author} {\bibfnamefont {Y.~S.}\ \bibnamefont
			{Patil}}, \bibinfo {author} {\bibfnamefont {J.}~\bibnamefont {H{\"o}ller}},
		\bibinfo {author} {\bibfnamefont {P.~A.}\ \bibnamefont {Henry}}, \bibinfo
		{author} {\bibfnamefont {C.}~\bibnamefont {Guria}}, \bibinfo {author}
		{\bibfnamefont {Y.}~\bibnamefont {Zhang}}, \bibinfo {author} {\bibfnamefont
			{L.}~\bibnamefont {Jiang}}, \bibinfo {author} {\bibfnamefont
			{N.}~\bibnamefont {Kralj}}, \bibinfo {author} {\bibfnamefont
			{N.}~\bibnamefont {Read}}, \ and\ \bibinfo {author} {\bibfnamefont {J.~G.}\
			\bibnamefont {Harris}},\ }\href {\doibase
		https://doi.org/10.1038/s41586-022-04796-w} {\bibfield  {journal} {\bibinfo
			{journal} {Nature}\ }\textbf {\bibinfo {volume} {607}},\ \bibinfo {pages}
		{271} (\bibinfo {year} {2022})}\BibitemShut {NoStop}%
\end{thebibliography}

\end{document}


\title{Supplemental Material for "Exceptional swallowtail degeneracies in driven-dissipative quadrature squeezing"}
\author{Polina Blinova}
\thanks{These authors contributed equally to this work.}
\affiliation{Department of Physics, McGill University, 3600 rue University, Montreal, QC, H3A 2T8, Canada.}
\affiliation{School of Applied and Engineering Physics, Cornell University, Ithaca, NY, 14850, USA.}
\author{Evgeny Moiseev}
\thanks{These authors contributed equally to this work.}
\affiliation{Department of Physics, McGill University, 3600 rue University, Montreal, QC, H3A 2T8, Canada.}
\author{Kai Wang}
\email{k.wang@mcgill.ca}
\affiliation{Department of Physics, McGill University, 3600 rue University, Montreal, QC, H3A 2T8, Canada.}
\maketitle
\tableofcontents
\section{I. \hspace{2pt} Degeneracies in bosonic quadratic Hamiltonians}
In the main text, we consider a Hamiltonian with real parameters, which describes two modes of an optical resonator coupled through a beam-splitter type of interaction, with one mode undergoing single-mode squeezing. More generally, the same structure can be described by the Hamiltonian. 
\fla{
\hat{\mathcal{H}} =
\delta\omega_1 \hat{a}^{\dagger}_1 \hat{a}_1 + \delta\omega_2 \hat{a}^{\dagger}_2 \hat{a}_2 + i ( \underline{g} \hat{a}^{\dagger}_1 \hat{a}_2 - \underline{g}^{*} \hat{a}_1 \hat{a}^{\dagger}_2  )  
+ \frac{i}{2}\left(\underline{\xi}_1 \hat{a}^{\dagger}_1 \hat{a}^{\dagger}_1 - \underline{\xi}^{*}_1 \hat{a}_1 \hat{a}_1  \right),
\label{eq::Ham_simple}
}
where $\hat{a}_{1(2)}$ $(\hat{a}^{\dagger}_{1(2)})$ is the annihilation (creation) operator for mode 1(2). Mode 1(2) has a resonance frequency of $\delta\omega_{1(2)} \in \mathbb{R}$ defined with respect to a properly chosen reference frame. The two modes experience beam-splitter-type coupling $\underline{g}$ with rate $g$ and phase $\phi_g$. Mode 1 experiences a parametric gain (squeezing) with a complex-valued parameter $\underline{\xi}_1=\xi_{1}e^{-i\phi_{1}}$ with magnitude $\xi_{1}$ and phase $\phi_{1}$.

The Heisenberg-Langevin equations of motion for the creation and annihilation operators are representable in a matrix form~\cite{Walls2008}:
\fla{
\frac{d\mathbf{a}}{dt}  = \tilde{\mathcal{E}} \cdot \mathbf{a}(t) + \mathcal{K} \cdot \mathbf{a}_{\text{in}}(t), \label{eq::EOM}
}
where $\tilde{\mathcal{E}}$ is the dynamical matrix obtained from canonical commutation relations, in which we also include losses $\gamma_{1(2)}$ for the two modes respectively, $\mathbf{a} = 
 \begin{pmatrix}
\hat{a}_1 &
\hat{a}_2 &
\hat{a}^{\dagger}_1 &
\hat{a}^{\dagger}_2
\end{pmatrix}^T, \;  
\mathbf{a}_{\text{in}} = 
 \begin{pmatrix}
\hat{a}_{1\text{in}} &
\hat{a}_{2\text{in}} &
\hat{a}^{\dagger}_{1\text{in}} &
\hat{a}^{\dagger}_{2\text{in}} 
\end{pmatrix}^T$, and  $\mathcal{K} = \text{diag}(\sqrt{\gamma_1},\sqrt{\gamma_2},\sqrt{\gamma_1},\sqrt{\gamma_2}).$ 

More specifically, the closed system of equations takes form
\fla{
\frac{d}{dt} \begin{pmatrix}
    \hat{a}_1 \\
\hat{a}_2  \\
\hat{a}^{\dagger}_1 \\
\hat{a}^{\dagger}_2
\end{pmatrix} = 
\begin{pmatrix}
-\gamma_1/2 -i\delta \omega_1 & ge^{-i\phi_g} & \xi_1e^{-i\phi_1} &  0 \\
-ge^{i\phi_g} & - \gamma_2/2 - i\delta \omega_2 & 0 & 0 \\
\xi_1e^{i\phi_1} & 0 & -\gamma_1/2 + i\delta \omega_1 &  ge^{i\phi_g} \\
0 & 0 & -ge^{-i\phi_g} & -\gamma_2/2 + i \delta \omega_2 
\end{pmatrix}
\begin{pmatrix}
    \hat{a}_1 \\
\hat{a}_2  \\
\hat{a}^{\dagger}_1 \\
\hat{a}^{\dagger}_2
\end{pmatrix}
+ \begin{pmatrix}
\sqrt{\gamma_1}\hat{a}_{1\text{in}} \\
\sqrt{\gamma_2} \hat{a}_{2\text{in}} \\
\sqrt{\gamma_1}\hat{a}^{\dagger}_{1\text{in}} \\
\sqrt{\gamma_2} \hat{a}^{\dagger}_{2\text{in}} 
\end{pmatrix}. \label{eq::dyn_simple}
}

\subsection{A. \hspace{2pt} Characteristic Polynomial}
The parameters controlling the eigenvalues $\lambda$ of an arbitrary $N\times N$ matrix $\mathcal{M}$ are the $N-1$ complex coefficients of the characteristic polynomial of $\mathcal{M}$:
\fla{\prod_{j=1}^n(\lambda-\lambda_j)=\lambda^N+ \underbrace{a_{N-1}}_{-Tr(\mathcal{M})}\lambda^{N-1} + ...+\underbrace{a_0}_{(-1)^N Det(\mathcal{M})}. \label{eq::gen_poly}}
The coefficients in~\eqref{eq::gen_poly} can also be expressed with elementary symmetric polynomials~\cite{michael_spivak_comprehensive_1979}:
\begin{equation}
    \prod_{j=1}^n(\lambda-\lambda_j)=\lambda^n-e_1(\lambda_1,...,\lambda_n)\lambda^{n-1}+e_2(\lambda_1,...,\lambda_n)\lambda^{n-2}+...+(-1)^n e_n(\lambda_1,...,\lambda_n),
\end{equation}
where $e_k(\lambda_1,...\lambda_n)=\sum_{1\le j_1<j_2<...<j_k\le n} \lambda_{j_1}...\lambda_{j_k}$.
Every matrix can be made traceless by subtracting the identity matrix multiplied by a constant, which simplifies the expressions for $e_k(\lambda_1,...\lambda_n)$. Importantly, this operation only offsets all the eigenvalues by the same constant while keeping eigenvectors the same. Hence, without loss of generality, we continue with the traceless dynamical matrix 
$
\mathcal{E} = \tilde{\mathcal{E}} - \text{Tr}(\tilde{\mathcal{E}})/4
$:
\fla{
\mathcal{E}=\left(
\begin{array}{cccc}
 -\frac{\gamma _-}{4}-i \delta\omega_1 & g e^{-i \phi _g} & \xi _1 e^{-i \phi_{1}} & 0 \\
  -ge^{i \phi _g} & \frac{\gamma _-}{4}-i\delta\omega_2 & 0 & 0 \\
 \xi _1 e^{i \phi_{1}} & 0 & -\frac{\gamma _-}{4}+i \delta\omega_1 & g e^{i \phi _g} \\
 0 & 0 & - g e^{-i \phi _g} & \frac{\gamma _-}{4}+i\delta\omega_2 \\
\end{array}
\right), \label{eq::dyn_matrix_simple}
}
where $\gamma_{\pm} = \gamma_1 \pm \gamma_2$.

We then use the coefficients of the characteristic polynomial $p(\lambda)$ of $\mathcal{E}$, defined as
\begin{equation}
p(\lambda)=\lambda^4+q\lambda^2 +r \lambda+s, \qquad \{q,r,s\} \in \mathbb{R} \label{eq::char},
\end{equation}
where $q=-\frac{1}{2}\text{Tr}\left[\mathcal{E}^2\right]$, $r=-\frac{1}{3}\text{Tr}\left[\mathcal{E}^3 \right]$, and $s=\text{det}\left[\mathcal{E} \right]$, to characterize the control parameter space and describe the degeneracies. In the absence of losses, the linear coefficient is zero, i.e., $r=0$ when $\gamma_{1,2}=0$.  The eigenvalue degeneracies are then the zero locus of the discriminant $D(q,r,s)$ of the characteristic polynomial in~\eqref{eq::char}:
\fla{D(q,r,s)=\prod_{i<j} (\lambda_i-\lambda_j)^2=16 q^4 s-4 q^3 r^2-128 q^2 s^2+144 q r^2 s-27 r^4+256 s^3 \label{eq::disc}.}
\subsection{B. \hspace{2pt} Symmetry of the dynamical matrix} 
We now characterize symmetries in the dynamical matrix $\mathcal{E}$, and by extension in $\tilde{\mathcal{E}}$. Since the identity matrix is invariant under all similarity transforms, the symmetries discussed below hold for both $\mathcal{E}$ and $\tilde{\mathcal{E}}$. The (traceless) dynamical matrix can generically be written in the block form
\fla{\mathcal{E}=\begin{pmatrix}
    \mathcal{A} & \mathcal{B} \\ \mathcal{C} & \mathcal{D}
\end{pmatrix},}
where $\{\mathcal{A}, \mathcal{B}, \mathcal{C}, \mathcal{D}\}$ are the $2\times2$ submatrices. The dynamical matrix exhibits particle-hole symmetry when $\mathcal{E}$ and $\mathcal{E}^*$ are similar with respect to $\tau_x=\sigma_x \otimes I_2$, i.e.,
\fla{\mathcal{E} = \tau_x \mathcal{E}^* \tau_x  \label{eq::PHS},}
where $\tau_j=\sigma_j \otimes I_2$, with $\sigma_j$ ($j=x,y,z$) being a Pauli matrix, $I_2$ being the $2\times 2$ identity matrix, and $\otimes$ denoting the Kronecker product. Particle-hole symmetry imposes the following constraints on the submatrices of $\mathcal{E}$:
\fla{\mathcal{A} = \mathcal{D}^*, \qquad \mathcal{B} = \mathcal{C}^* \label{eq::phscond}.}
If the Hamiltonian is Hermitian, the dynamical matrix also exhibits pseudo-Hermiticity
\fla{\mathcal{E} = -\tau_z \mathcal{E}^\dag \tau_z \label{eq::PH},}
defined using $\tau_z = \sigma_z \otimes I$. We note that the transformation $\mathcal{E}\rightarrow i \mathcal{E}$ recovers a more familiar definition of pseudo-Hermiticity without the negative sign~\cite{melkani}. We also emphasize that particle-hole symmetry is in a broader mathematical sense also a version of pseudo-Hermiticity~\cite{melkani}. We use "particle-hole symmetry" to describe similarity with respect to $\tau_x$ and limit "pseudo-Hermiticity" to similarity with respect to $\tau_z$. Pseudo-Hermiticity dictates the following relations on the submatrices of $\mathcal{E}$:
\fla{\mathcal{A} = -\mathcal{A}^\dag, \qquad \mathcal{B} = \mathcal{C}^\dag, \qquad \mathcal{D} = -\mathcal{D}^\dag \label{eq::phcond}.}
In the presence of both symmetries, the constraints~\eqref{eq::phscond} and~\eqref{eq::phcond} in addition dictate that
\fla{\mathcal{B}=\mathcal{B}^T, \qquad \mathcal{C} = \mathcal{C}^T,}
and $\mathcal{E}$ takes on the form
\fla{\mathcal{E} = \begin{pmatrix}
    \mathcal{A} & \mathcal{B} \\ \mathcal{B}^* & \mathcal{A}^*
\end{pmatrix}=\begin{pmatrix}
    \mathcal{A} & \mathcal{B} \\ \mathcal{B}^* & -\mathcal{A}^T
\end{pmatrix}.}

The particle-hole symmetry consequences on eigenvalues of $\mathcal{E}$ are as follows: Consider first an arbitrary right eigenvector $|\textbf{R}\rangle $ of $\mathcal{E}$ with its eigenvalue $\lambda$, i.e.,
\fla{\mathcal{E}|\textbf{R}\rangle = \lambda |\textbf{R}\rangle \label{eigv}.}
The condition~\eqref{eigv} is then reformulated using~\eqref{eq::PHS} as
\fla{\left(\tau_x \mathcal{E}^* \tau_x\right) |\textbf{R}\rangle = \lambda |\textbf{R}\rangle\quad \rightarrow \quad \mathcal{E}^* \left(\tau_x |\textbf{R}\rangle \right) =\lambda (\tau_x|\textbf{R}\rangle),}
and conjugating the above equation gives
\fla{\mathcal{E} (\tau_x |\textbf{R}\rangle)^* = \lambda^* (\tau_x |\textbf{R}\rangle)^* \label{phsym}.}
It follows that a complex eigenvalue of $\mathcal{E}$ must always come in a complex-conjugate pair. 
Particle-hole symmetry also implies that if $|\textbf{R}\rangle$ is an eigenvector of $\mathcal{E}$ with eigenvalue $\lambda$, then there exists an eigenvector $\tau_x |\textbf{R}\rangle^*$ of $\mathcal{E}$ with eigenvalue $\lambda^*$. This does not necessarily mean that the real eigenvalues of $\mathcal{E}$ are degenerate. 
Particle-hole symmetry states that $\lambda\in\mathbb{R}$ if and only if $|\textbf{R}\rangle$ follows the linear dependence relationship $|\textbf{R}\rangle=\eta\tau_x|\textbf{R}\rangle^*$ for some constant $\eta$ ~\cite{melkani}. 

For the dynamical matrix derived from Hamiltonian~\eqref{eq::Ham_simple}, particle-hole symmetry ensures that the coefficients of $p(\lambda)$ are real. Therefore, the largest subspace spanned by $\{q,r,s\}$ for a general two mode dynamical matrix is $\mathbb{R}^3$. Since any given eigenvalue of $\mathcal{E}$ is either real or occurs in a complex-conjugate pair, the reality of $\{q,r,s\}$ is easily proven. In the case when $\lambda_1=\lambda_2^*, \lambda_3=\lambda_4^*$, the coefficients reduce to 
\begin{align*}
    q=e_2&=\lambda_1 \lambda_2+\lambda_1 \lambda_3+\lambda_1 \lambda_4+\lambda_2 \lambda_3+\lambda_2 \lambda_4+\lambda_3 \lambda_4\\
    &=|\lambda_1|^2+|\lambda_3|^2+\lambda_1\lambda_3+\lambda_1^*\lambda_3+\text{c.c} \; \in \mathbb{R}, \\
    r=e_3&=\lambda_1 \lambda_2 \lambda_3+\lambda_1 \lambda_2 \lambda_4+\lambda_1 \lambda_3 \lambda_4+\lambda_2 \lambda_3 \lambda_4\\
    &=|\lambda_1|^2\lambda_3+|\lambda_3|^2\lambda_1+\text{c.c.} \; \in \mathbb{R}, \\
    s=e_4&=\lambda_1 \lambda_2 \lambda_3 \lambda_4=|\lambda_1|^2|\lambda_3|^2 \; \in \mathbb{R}.
\end{align*}
If eigenvalues come as $\lambda_1=\lambda_2^*,\; \{\lambda_3,\lambda_4\}\in\mathbb{R}$, then
\begin{align*}
    q=e_2&=|\lambda_1|^2+(\lambda_1\lambda_3+\text{c.c})+(\lambda_1\lambda_4+\text{c.c})+\lambda_3\lambda_4\ \; \in \mathbb{R}, \\
    r=e_3&=|\lambda_1|^2\lambda_3+|\lambda_1|^2\lambda_4+(\lambda_1\lambda_3\lambda_4+\text{c.c}) \; \in \mathbb{R}, \\
    s=e_4&=|\lambda_1|^2\lambda_3\lambda_4 \; \in \mathbb{R}.
\end{align*}
In the trivial case when all the eigenvalues are real, the coefficients of $p(\lambda)$ are also real. Likewise, the consequences of pseudo-Hermiticity follow from~\eqref{eq::PH}, i.e.,
\fla{(\tau_z \mathcal{E}^\dag \tau_z) |\textbf{R}\rangle = -\lambda |\textbf{R}\rangle \quad \rightarrow \quad \mathcal{E}^\dag (\tau_z |\textbf{R}\rangle) =-\lambda (\tau_z|\textbf{R}\rangle),}
and conjugating the above equation gives
\fla{\mathcal{E}^T(\tau_z|\textbf{R}\rangle^*) = -\lambda^* (\tau_z
|\textbf{R}\rangle^*)\label{pseudosym}.}
Since any matrix and its transpose share eigenvalues, it follows that a complex eigenvalue $\lambda$ occurs together with $-\lambda^*$. For a Hermitian Hamiltonian, both particle-hole symmetry and pseudo-Hermiticity hold, and any eigenvalue  $\lambda$ is either real or occurs in a quartet $\{\lambda,-\lambda,\lambda^*,-\lambda^*\}$. The form of $\mathcal{E}$ in the presence of both symmetries restricts the control space to $\mathbb{R}^2$, which follows from calculating $r=-\frac{1}{3}\text{Tr}\left[\mathcal{E}^3\right]$. First, we have
\begin{equation}
\begin{split}
    \mathcal{E}^3&=\begin{pmatrix}
        \mathcal{A}&\mathcal{B}\\\mathcal{B}^*&\mathcal{A}^*
    \end{pmatrix}^3 = \begin{pmatrix}
        \mathcal{A} & \mathcal{B} \\ \mathcal{B}^* & -\mathcal{A}^T
    \end{pmatrix}^3 \\
    &=\begin{pmatrix}
    \mathcal{A}^3+\mathcal{A}\mathcal{B}\mathcal{B}^*+\mathcal{B}\mathcal{B}^*\mathcal{A}-\mathcal{B}\mathcal{A}^T\mathcal{B}^*&...\\ ...&\mathcal{B}^*\mathcal{A}\mathcal{B}-\mathcal{B}^*\mathcal{B}\mathcal{A}^T-\mathcal{A}^T\mathcal{B}^*\mathcal{B}-(\mathcal{A}^T)^3
    \end{pmatrix}.
\end{split}
\end{equation}
Calculating the trace gives
\begin{equation}\begin{split}
    \text{Tr}\left[\mathcal{E}^3\right]&= \text{Tr}\left[\mathcal{A}^3\right]- \text{Tr}\left[(\mathcal{A}^T)^3\right]+3\text{Tr}\left[\mathcal{A}\mathcal{B}\mathcal{B}^*\right]-3 \text{Tr}\left[\mathcal{A}^T\mathcal{B}^*\mathcal{B}\right] \\
    &=3 \text{Tr}\left[(\mathcal{A}\mathcal{B}\mathcal{B}^*)^T\right]-3 \text{Tr}\left[\mathcal{A}^T\mathcal{B}^*\mathcal{B}\right]=\text{Tr}\left[\mathcal{A}^T(\mathcal{B}^\dag\mathcal{B}^T-\mathcal{B}^*\mathcal{B} )\right]=0,
\end{split}\end{equation}
which is zero because $\mathcal{B}$ is a symmetric matrix. In fact, the above would hold even if the off-diagonal blocks $\mathcal{B}$ and $\mathcal{C}$ were distinct symmetric matrices.

\subsection{C. \hspace{2pt} More general model}
Swallowtail degeneracy extends beyond two-mode Hamiltonians with single-mode squeezing and beam-splitter interactions, as mentioned in the main text. In general, we may consider two bosonic modes that are governed by an arbitrary quadratic Hamiltonian:
\fla{
\hat{\mathcal{H}} =
\delta\omega_1 \hat{a}^{\dagger}_1 \hat{a}_1 + \delta\omega_2 \hat{a}^{\dagger}_2 \hat{a}_2 + i ( \underline{g} \hat{a}^{\dagger}_1 \hat{a}_2 - \underline{g}^{*} \hat{a}_1 \hat{a}^{\dagger}_2  ) +i (\underline{\chi} \hat{a}^{\dagger}_1 \hat{a}^{\dagger}_2 - \underline{\chi}^{*} \hat{a}_2 \hat{a}_1 ) +  \nonumber \\
+ \frac{i}{2}\left(\underline{\xi}_1 \hat{a}^{\dagger}_1 \hat{a}^{\dagger}_1 - \underline{\xi}^{*}_1 \hat{a}_1 \hat{a}_1  \right)+ \frac{i}{2} \left(\underline{\xi}_2 \hat{a}^{\dagger}_2 \hat{a}^{\dagger}_2 - \underline{\xi}^{*}_2  \hat{a}_2 \hat{a}_2  \right),
\label{eq::Ham}
}
where we add two complex-valued parameters to the original Hamiltonian~\eqref{eq::Ham_simple}: two-mode squeezing gain $\underline{\chi}=\chi e^{ -i\phi_{\chi}}$ with magnitude $\chi$ and phase $\phi_{\chi}$; single-mode squeezing gain $\underline{\xi}_2=\xi_{2}e^{-i\phi_{2}}$ for mode 2 with magnitude $\xi_2$ and phase $\phi_{2}$.
The dynamical matrix $\tilde{\mathcal{E}}'$ is now:
\fla{
\tilde{\mathcal{E}}' = 
\begin{pmatrix}
-\gamma_1/2 -i\delta \omega_1 & ge^{-i\phi_g} & \xi_1e^{-i\phi_1} &  \chi e^{-i\phi_\chi} \\
-ge^{i\phi_g} & - \gamma_2/2 - i\delta \omega_2 & \chi e^{-i\phi_\chi}  & \xi_2 e^{-i\phi_2} \\
\xi_1 e^{i\phi_1} & \chi e^{i\phi_\chi} & -\gamma_1/2 + i\delta \omega_1 &  g e^{i\phi_g} \\
\chi e^{i\phi_\chi} & \xi_2 e^{i\phi_2} & -ge^{-i\phi_g} & -\gamma_2/2 + i \delta \omega_2 
\end{pmatrix}.
\label{eq::dyn_general}
}
The conclusions about eigenvalue degeneracies above hold for dynamical matrices derived from Hamiltonians ~\eqref{eq::Ham_simple} and~\eqref{eq::Ham}, since the block matrix form of $\tilde{\mathcal{E}}$ and $\tilde{\mathcal{E}}'$ is identical. In particular, the swallowtail degeneracy is a common feature of bosonic quadratic Hamiltonians.

\section{II. \hspace{2pt} Swallowtail Degeneracy}
\begin{figure}[!hbt]
    \centering
    \includegraphics[width=.7\textwidth]{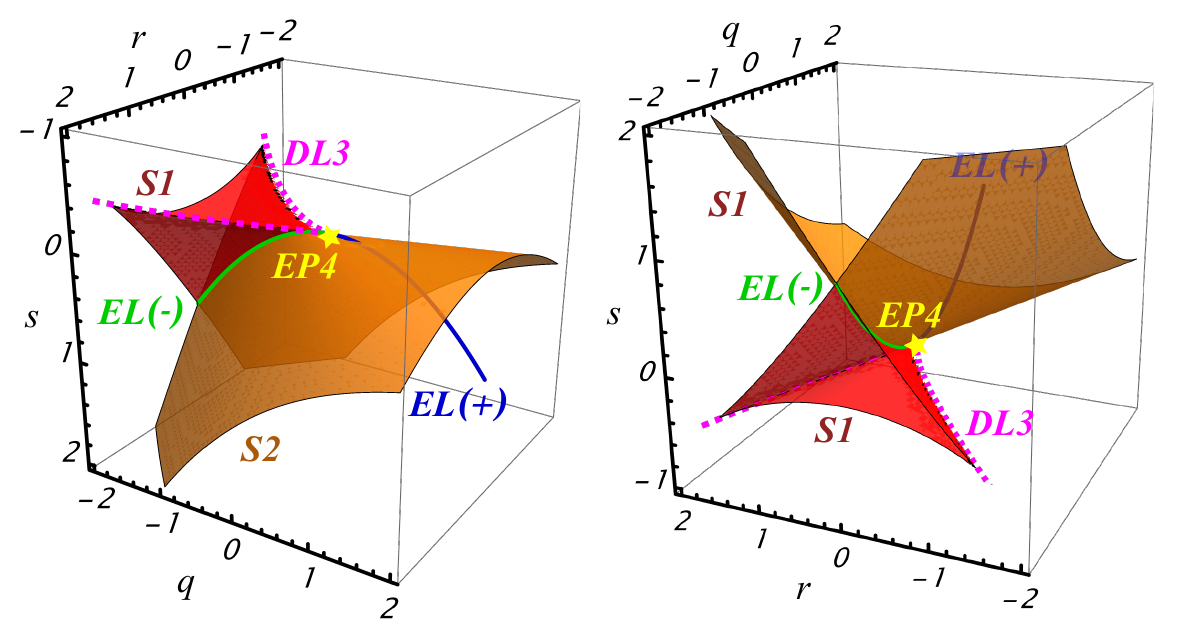}
    \caption{Swallowtail degeneracy viewed from different angles. The degeneracy structure features: a fourth-order exceptional point EP4 (yellow star); a line of pairwise two-fold exceptional degeneracies EL, separated into real (green) and imaginary (blue) eigenvalues on EL($-$) and EL($+$) respectively; and a line of three-fold diabolical degeneracies DL3 (pink, dashed). S1 features all real eigenvalues, while there are pairs of real and complex-conjugate eigenvalues on S2.}
    \label{fig:sw_alone}
\end{figure}

The eigenvalue degeneracy structure of dynamical matrices in~\eqref{eq::dyn_simple} and~\eqref{eq::dyn_general} is the set of zeros of the discriminant~\eqref{eq::disc} of the depressed quartic polynomial~\eqref{eq::char}. Formally, the discriminant variety of a $4^{\text{th}}$ order polynomial is ambient diffeomorphic to a direct product of a line \(\mathbb{R}^1\) and the "swallowtail"~\cite{vassiliev1995topology}, where the latter is a known catastrophe in Arnold's ADE classification. More on ADE classification can be found in Ref.~\cite{batorova2013desingularization}.
The topology of the swallowtail is non-trivial, as it hosts higher-order degeneracies, which are described in the main text. The degeneracy structure is depicted on Fig. \ref{fig:sw_alone}, as viewed from different angles. It is formed in the control space spanned by coefficients 
\fla{
q &= 2 \left(g^2 - \frac{\gamma^{2}_-}{16} \right) - (\xi_1^2-\delta\omega_1^2)+\delta\omega_2^2, \label{eq:q}\\
r &= \frac{\gamma_-}{2} (\xi_1^2-\delta\omega_1^2+\delta\omega_2^2), \label{eq:r}\\
s &= \left(g^2 - \frac{\gamma^{2}_-}{16} \right)^2 - \frac{\gamma^{2}_-}{16} (\xi_1^2-\delta\omega_1^2-\delta\omega_2^2)-2g^2\delta\omega_1\delta\omega_2-(\xi_1^2-\delta\omega_1^2)\delta\omega_2^2
\label{eq:s}}
in model derived from dynamical matrix ~\eqref{eq::dyn_matrix_simple}. In the more general dynamical matrix~\eqref{eq::dyn_general}, the coefficients are
\fla{
q &= -\frac{\gamma _-^2}{8}+2 g^2-\xi _1^2-\xi _2^2-2 \chi ^2+\delta\omega _1^2+\delta\omega _2^2, \\
r &= \frac{1}{2} \gamma _- \left(\xi _1^2-\xi _2^2-\delta\omega _1^2+\delta\omega _2^2\right), \\
s &= \frac{1}{16} \gamma_{-}^2 \left(2 \chi ^2+\delta\omega _1^2+\delta\omega _2^2 -\xi _1^2-\xi _2^2\right)+\frac{\gamma _-^4}{256}+g^4-\frac{1}{8} g^2 \left(\gamma _-^2+16 \left(\chi ^2+\delta\omega _1 \delta\omega _2\right)\right) +
\nonumber \\
&+4 g \chi  \left(\xi _2 \delta\omega _1 \sin \left(\phi _g-\phi _{\chi }+\phi_{2}\right)-\xi _1 \delta\omega _2 \sin \left(-\phi _g-\phi _{\chi }+\phi_{1}\right)\right)+(\xi _2^2 -\delta\omega^{2}_2) \left(\xi^{2}_1-\delta\omega^{2}_1\right) 
\nonumber \\
&2 \xi _1 \xi _2  \left( 
g^2 \cos \left(-2 \phi _g+\phi_{1}-\phi_{2}\right)
-\chi ^2 \cos \left(-2 \phi _{\chi }+\phi_{1}+\phi_{2}\right) 
\right)
+\chi ^4-2 \chi ^2 \delta\omega_1 \delta\omega_2. 
}
\subsection{A. \hspace{2pt} Mapping the Hamiltonian to the control parameter space}
Our model~\eqref{eq::Ham_simple} allows for an analytical mapping between the coefficients of $p(\lambda)$ and $\hat{\mathcal{H}}$ when $\delta\omega_2=0$. We can use inverse function theorem to argue that the parameters in the characteristic polynomial are invertible to the bare control parameters in the Hamiltonian under some condition. We define the Jacobian of a mapping $\boldsymbol{f}: (q,r,s)\rightarrow (\gamma_{-},\xi_1,g)$, treating $\delta\omega_{1,2}$ as constants:
\fla{J_f &= \begin{pmatrix}
    \frac{\partial q}{\partial \gamma} & \frac{\partial q}{\partial \xi_1} & \frac{\partial q}{\partial g} \\[4pt]
    \frac{\partial r}{\partial \gamma} & \frac{\partial r}{\partial \xi_1} & \frac{\partial r}{\partial g} \\[4pt]
    \frac{\partial s}{\partial \gamma} & \frac{\partial s}{\partial \xi_1} & \frac{\partial s}{\partial g} 
\end{pmatrix} = \nonumber \\
&\begin{pmatrix} 
    -\frac{\gamma_{-}}{4} & -2\xi_1 & 4g \\[5pt]
    \frac{1}{2}\left(\xi_1^2-\delta\omega_1^2+\delta\omega_2^2\right) & \gamma_{-}\xi_1 & 0\\[5pt]
    \frac{1}{64}\gamma_{-}\left[\gamma_{-}^2-16g^2+8\left(\delta\omega_1^2+\delta\omega_2^2-\xi_1^2\right)\right] & -\frac{1}{8}\xi_1\left(\gamma_{-}^2+16\delta\omega_2^2\right) & 4g^3 - \frac{g}{4}\left(\gamma_{-}^2+16\delta\omega_1\delta\omega_2\right)
 \end{pmatrix},}
By inverse function theorem, a continuously differentiable function $\boldsymbol{f}$ is locally invertible in the neighbourhood of a given point $\boldsymbol{x} = (q,r,s)$ \textit{iff} the Jacobian determinant is non-zero~\cite[Theorem~2-11]{spivak2018calculus}. In our case, the Jacobian determinant is
\fla{\text{Det}(J_f) = g\xi_1\left\{4g^2\left[\xi_1^2-\delta\omega_1^2+\delta\omega_2^2\right]-\delta\omega_2\left[\gamma_{-}^2(\delta\omega_2-\delta\omega_1)+4(\delta\omega_1+\delta\omega_2)(\xi_1^2-\delta\omega_1^2+\delta\omega_2^2) \right] \right \},}
which has a more illuminating form when $\delta\omega_2=0$ (as in Fig. 3(a) in the main text):
\fla{\text{Det}(J_f) = 4g^3\xi_1\left(\xi_1^2-\delta\omega_1^2\right).}
This quantity is non-zero as long as \textbf{none of $\boldsymbol{g, \xi_1, \xi_1^2-\delta\omega_1^2}$ are zero}. If this condition is respected, then there is a well-defined inverse mapping at that point in parameter space. If $\delta\omega_2\ne 0$, the invertibility condition is more complicated, in particular,
\fla{g\ne0, \qquad \xi_1 \ne0, \qquad \gamma_{-}\ne\pm \frac{2i\sqrt{g^2-\delta\omega_1\delta\omega_2-\delta\omega_2^2}\sqrt{\xi_1^2-\delta\omega_1^2+\delta\omega_2^2}}{\sqrt{\delta\omega_1\delta\omega_2-\delta\omega_2^2}}.}
\begin{figure}[!htb]
    \centering
    \includegraphics[width=0.4\linewidth]{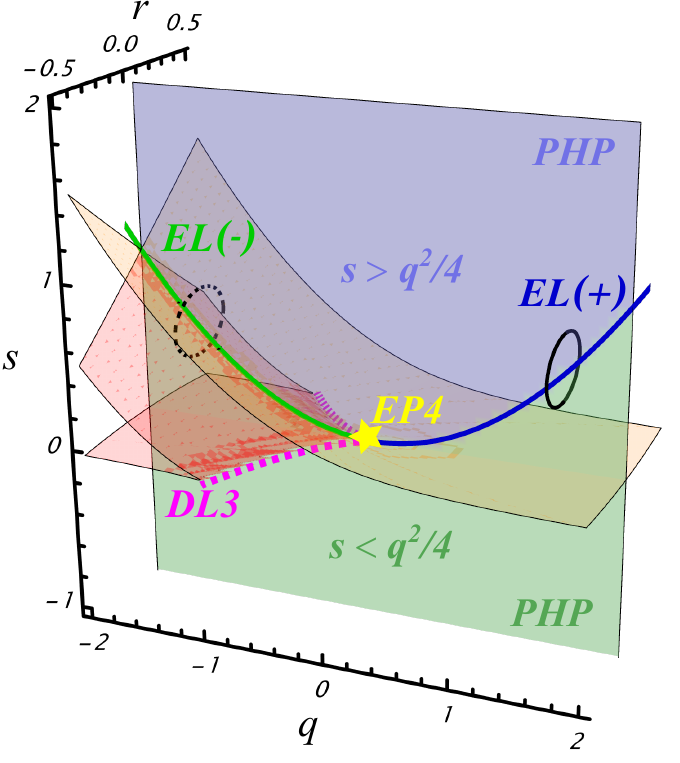}
    \caption{Parametric loops around the EL. To realize a loop around the EL, it is necessary to cross PHP once for $s>q^2/4$ (blue area) and once for $s<q^2/4$ (green area). Solid black loop around EL($+$) is only possible with $\delta\omega_2 \ne 0$. Dashed black loop is possible with $\delta \omega_2=0$.}
    \label{fig:omega2}
\end{figure}

The limitation of the special case with $\delta\omega_2=0$ is the inability to realize a loop around EL($-$) to achieve eigenvalue braids. We can understand it as follows: We must traverse the PHP twice to loop around EL($-$), and we achieve this by setting $r=0$ via
\fla{\xi_1^2-\delta\omega_1^2=0 \qquad \text{or} \qquad \gamma_{-}=0.}
Under these conditions, the explicit form of the mapping equation~\eqref{eq:s} is
\fla{s=\frac{q^2}{4}-2g\delta\omega_1\delta\omega_2+g^2\left(\xi_1^2-\delta\omega_1^2-\delta\omega_2^2\right)-\frac{1}{4}\left(\xi_1^2-\delta\omega_1^2+\delta\omega_2^2\right)^2 \label{eq::gammazero},}
for $\gamma_{-}=0$, or
\fla{s=\frac{q^2}{4}+\frac{\delta\omega_2}{4}\left(\gamma_{-}^2\delta\omega_2-8g\delta\omega_1-8g^2\delta\omega_2\right), \label{eq::uzero}}
for $\xi_1^2-\delta\omega_1^2=0$. We readily see that if $\delta\omega_2=0$, then crossing the PHP via~\eqref{eq::uzero} is impossible without going through the EL ($s=q^2/4, \; r=0$). If conversely, we set $\gamma_{-}=0$ with $\delta\omega_2=0$, then
\fla{q=2g^2-\left(\xi_1^2-\delta\omega_1^2\right), \qquad s=g^4.\label{eq::qEL}}
To complete a loop around the EL, the trajectory needs to cross the PHP once for $s>q^2/4$ and once for $s<q^2/4$ (see Fig.~\ref{fig:omega2}), which implies going through the point $s=\frac{q^2}{4}$ outside of the PHP. In bare Hamiltonian parameters, this condition corresponds to $\xi_1^2-\delta\omega_1^2=4g^2$, which reduces~\eqref{eq::qEL} to $q=-2g^2$. The equation for $q$ is strictly negative, therefore, only loops around EL($-$) are possible (dashed loop on Fig.~\ref{fig:omega2}), and not EL($+$) (solid loop on Fig.~\ref{fig:omega2}). The trajectory must not touch the degeneracy to realize eigenvalue braids, but EL($-$) is situated on the swallowtail. Therefore, it is necessary to include a non-zero detuning on the second mode to achieve eigenvalue braids around EL($+$). 

\subsection{B. \hspace{2pt} Numerically spanning the control parameter space}
To numerically reconstruct the swallowtail and form various loops in control parameter space as in Fig. 3 of the main text, we set the phases of all complex-valued parameters in the Hamiltonian~\eqref{eq::Ham_simple} to zero and vary the amplitudes. To form loops in control parameter space, we introduce parameter modulation:
\fla{
\xi_1 \rightarrow a_{\xi}\left( 1+m_{\xi} \text{cos}\phi\right), \\
g \rightarrow a_{g}\left( 1+m_{g} \text{sin}\phi\right), \\
\gamma_{-} \rightarrow a_{\gamma}\left( 1+m_{\gamma} \text{cos}\phi\right),
}
where $a_{\xi,g,\gamma}$ and $m_{\xi,g,\gamma}$ are real-valued. In Fig. 3(b) of the main text we use $a_{\xi}=1.5, \; a_{g} = 1.5, \; a_{\gamma}=1, \; m_{\xi}=0.01, \; m_g=0.01, \; m_{\gamma}=0.1$, and in Fig. 3(c) of the main text we use $a_{\xi}=1.5, \; a_{g} = 1.4, \; a_{\gamma}=0.1, \; m_{\xi}=0.1, \; m_g=0.1, \; m_{\gamma}=2$. We also set the frequencies $\delta\omega_1=0, \; \delta\omega_2=0$ to form a trivial braid, and $\delta\omega_1=0, \; \delta\omega_2=0.92$ to form a nontrivial one.
\subsection{C. \hspace{2pt} Swallowtail Degeneracy Classification}
Based on the classification of roots of a $4^{\text{th}}$ order depressed quartic polynomial~\cite{ARNON198837}, we provide a full classification of swallowtail degeneracies in terms of our model~\eqref{eq::Ham_simple}. For simplicity, we set $\delta\omega_2=0$ and define $u\equiv\xi_1^2-\delta\omega_1^2$. The classification is based on forming the following quantity, called the cubic resolvent:
\fla{L = -2 q^3 - 9 r^2 + 8 q s,}
or in Hamiltonian parameters
\fla{L=\frac{1}{8} \left[128 g^4 + (\gamma_{-}^2 - 4 u )^2 - 24 g^2 (\gamma_{-}^2 + 4 u ) \right],}
with $u = \xi^{2}_1 - \delta \omega^{2}_1 $.
The detailed classification, based on Ref.~\cite{ARNON198837} is provided in Table~\ref{tab::degen}. 
\begin{table}[!hbt]
\renewcommand{\arraystretch}{3.5}
\centering
\begin{tabular}{|p{1.2cm}|p{3.2cm}|p{3.5cm}|p{4cm}|p{1.75cm}|p{1cm}| p{2.2cm}| }
\hline
 & Condition & Eigenvalues & Equation in $u,g,\gamma_{-}$& q & r & s \\
\hline
EL(--) & $L=0$, $q<0$, $r = 0$ & $+\frac{i\sqrt{q}}{\sqrt{2}}$ (2) $\quad  -\frac{i\sqrt{q}}{\sqrt{2}} (2)$&  \shortstack[l]{$u=0  \; (\text{\tiny EL}^{\textbf{(I)}})$   \\ $\gamma_{-}=0,u=4g^2 \; (\text{\tiny EL}^{\textbf{(II)}})$} & \shortstack[l]{$2g^2-\gamma_{-}^2/8 $\\ $-2g^2$} & $0$ & \shortstack[l]{$\left(g^2-\gamma_{-}^2/16\right)^2$ \\ $g^4$} \\
\hline
EL($+$) & $L=0$, $q>0$, $r = 0$ & $+\frac{i\sqrt{q}}{\sqrt{2}}$ (2) $\quad  -\frac{i\sqrt{q}}{\sqrt{2}} (2)$ & $u=0 \;  (\text{\tiny EL}^{\textbf{(I)}})$ & $2g^2-\gamma_{-}^2/8 $ & $0$ & $\left(g^2-\gamma_{-}^2/16\right)^2$ \\
\hline
DL3 & $L=0$, $q<0$, $r \ne 0$ & $\pm i \frac{\sqrt{3q}}{\sqrt{2}}$ (1), $\; \mp i\frac{\sqrt{q}}{6}$ (3) & $g=0,u=\frac{\gamma_{-}^2}{4}$ & $-\frac{3\gamma_{-}^2}{8}$ & $\frac{\gamma_{-}^3}{8}$ & $-\frac{3 \gamma_{-}^4}{256}$ \\
\hline
EP4 & $L=0$, $q=0$ & $0$ (4) &\shortstack[l]{$u=0, \gamma_{-}=\pm 4g\; {(\text{\tiny EP}^{\textbf{(I)}})}$ \\ $u,g,\gamma_{-}=0\; {(\text{\tiny EP}^{\textbf{(II)}})}$}  & $0$ & $0$& $0$\\
\hline
\end{tabular}
\caption{Degeneracy classification on the swallowtail.  Provided are: general condition in terms of $L$ and coefficients (column 2); expressions for eigenvalues including algebraic multiplicity in brackets (column 3); equation in bare Hamiltonian parameters (column 4), as well as coefficients of $p(\lambda)$ themselves (columns 5-7). We set $\delta\omega_2=0$ in this classification.}
\label{tab::degen}
\end{table}
To characterize whether a degeneracy is exceptional or not, we form a matrix $\mathcal{U}$ out of eigenvectors $|\textbf{R}_i\rangle$ of $\mathcal{E}$:
\fla{
\mathcal{U} = \begin{pmatrix} |\textbf{R}_1\rangle &|\textbf{R}_2\rangle & |\textbf{R}_3\rangle &|\textbf{R}_4\rangle \end{pmatrix}
\label{eq::eig-vec-matr}
}
and calculate its determinant at given point.
If $\text{det} \mathcal{U} = 0$, eigenvectors are linearly dependent, and the degeneracy at that point is exceptional. We provide a more detailed characterization of the degeneracies below, where we set $\gamma_{+}=0$ that just shifts all eigenvalues equally along the real axis.

\subsubsection{1. \hspace{2pt} Pairs of two-fold defective degeneracies EL($\pm$)}

In our model~\eqref{eq::Ham_simple}, one way to achieve EL is by setting $u=0$ (Table~\ref{tab::degen}, $\text{EL}^{\textbf{(I)}}$). We find that $\mathcal{E}$ is defective along the whole line, and for each degenerate eigenvalue, there is a pair of degenerate eigenvectors:
\fla{|\textbf{R}_{1,2}^{\text{EL}}\rangle = \left( -\frac{ e^{i(\phi_g-\phi_{1})}}{4g}  \left(i\gamma_{-}+\sqrt{16g^2-\gamma_{-}^2}\right), \; -i e^{2i\phi_g-i\phi_{1}}, \; \frac{ e^{i\phi_g}}{4g}  \left(\gamma_{-}-i\sqrt{16g^2-\gamma_{-}^2}\right), 1 \right)^T, \\
|\textbf{R}_{3,4}^{\text{EL}}\rangle = \left( \frac{e^{i(\phi_g-\phi_{1})}}{4g}  \left(-i\gamma_{-}+\sqrt{16g^2-\gamma_{-}^2}\right), \; -i e^{2i\phi_g-i\phi_{1}}, \; \frac{e^{i\phi_g}}{4g}  \left(\gamma_{-}+i\sqrt{16g^2-\gamma_{-}^2}\right), 1 \right)^T.}
Another way to achieve EL($-$) is with $\gamma_{-}=0, u=4g^2$ (Table~\ref{tab::degen}, $\text{EL}^{\textbf{(II)}}$). In this case, the degenerate eigenvectors of $\mathcal{E}$ are
\fla{|\textbf{R}_{1,2}^{\text{EL}}\rangle=\left( -\frac{e^{i(\phi_g-\phi_{1})}}{\xi_1} \left(2g+i\sqrt{-4g^2+\xi_1^2}\right),-\frac{e^{i(2\phi_g-\phi_{1})}}{\xi_1}\left(2g+i\sqrt{-4g^2+\xi_1^2}\right),e^{i\phi_g},1 \right)^T, \\
|\textbf{R}_{3,4}^{\text{EL}}\rangle = \left( \frac{e^{i(\phi_g-\phi_{1})}}{\xi_1}\left( -2g+i\sqrt{-4g^2+\xi_1^2}\right), \frac{e^{i(2\phi_g-\phi_{1})}}{\xi_1}\left( 2g -i\sqrt{-4g^2+\xi_1^2}\right), -e^{i\phi_g},1\right)^T. }
\subsubsection{2. \hspace{2pt} Four-fold defective degeneracy EP4}

An obvious way to reach EP4 in our model is by setting all available parameters to zero (Table~\ref{tab::degen}, $\text{EP}^{\textbf{(I)}}$). Interestingly, although the eigenvalues are four-fold degenerate at the origin of control parameter space, there is only one pair of degenerate eigenvectors:
\fla{|\textbf{R}_1^{\text{EP4}}\rangle &=\left(0,0,0,1\right)^T, \\
|\textbf{R}_2^{\text{EP4}}\rangle &=\left(-ie^{-i\phi_{1}},0,1,0\right)^T, \\
|\textbf{R}_{3,4}^{\text{EP4}}\rangle &=\left(0,1,0,0\right)^T.}
However, using a non-trivial condition for EP4, $u=0, \gamma_{-}=\pm4g$ (Table~\ref{tab::degen}, $\text{EP}^{\textbf{(II)}}$), we obtain a four-fold degenerate eigenvector
\fla{|\textbf{R}_{1,2,3,4}^\text{EP4}\rangle = \left( \mp ie^{i(\phi_g-\phi_{1})},-ie^{i(2\phi_g-\phi_{1})},\pm e^{i\phi_g},1\right)^T.}

\subsubsection{3. \hspace{2pt} Three-fold non-defective degeneracy DL3}

The three-fold degeneracy can be achieved by setting $g=0,\;u=\gamma_{-}^2/4$. On these lines, $\mathcal{E}$ is always non defective in our model, and eigenvectors are
\fla{|\textbf{R}_1^{\text{DL3}}\rangle &=\left(0,0,0,1\right)^T, \\
|\textbf{R}_2^{\text{DL3}} \rangle &= \left( \frac{e^{-i\phi_{1}}}{2\xi_1} \left( \gamma_{-}-i\sqrt{-\gamma_{-}^2+4\xi_1^2}\right),0,1,0\right)^T,\\
|\textbf{R}_3^{\text{DL3}}\rangle & = \left(0,1,0,0 \right)^T, \\
|\textbf{R}_4^{\text{DL3}}\rangle &= \left( -\frac{e^{-i\phi_{1}}}{2\xi_1}\left(\gamma_{-}+i\sqrt{-\gamma_{-}^2+4\xi_1^2} \right),0,1,0\right)^T.}
By substituting the above eigenvectors in Eq. \eqref{eq::eig-vec-matr}, we get the matrix $\mathcal{U}^{\text{DL3}} $ with its determinant being
\fla{ 
\text{det} \mathcal{U}^{\text{DL3}} 
= -\frac{e^{-i\phi_{1}}\gamma_{-}}{\xi_1},}
which is non-zero outside of PHP. Zero value at PHP ($\gamma_{-}=0$) corresponds to the point at the origin (EP4).

\subsubsection{4. \hspace{2pt} Degenerate surfaces S1 and S2}
\begin{figure}[!htb]
    \centering
    \includegraphics[width=1\textwidth]{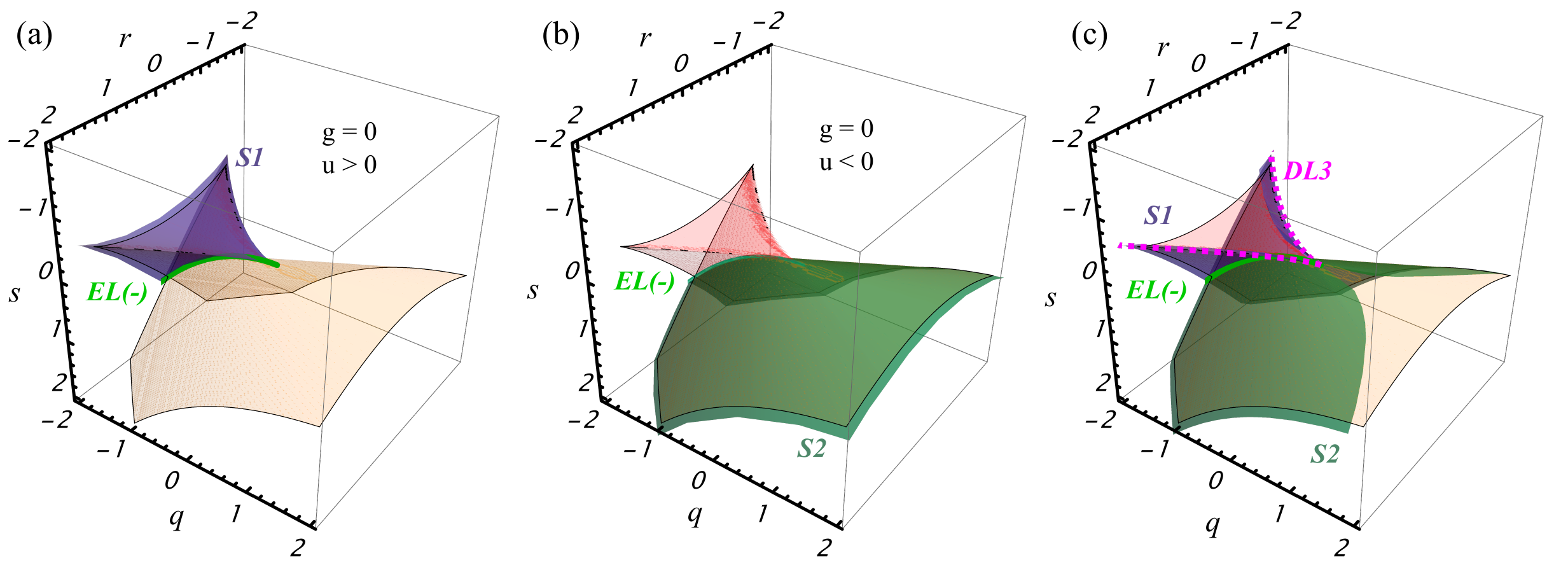}
    \caption{Mapping of different regions of the swallowtail. Purple (S1) and green (S2) are numerically reconstructed and plotted over the analytical swallowtail (opaque orange/red as in Fig.~\ref{fig:sw_alone}). S1 and S2 are diabolical when $g=0$, with S1 corresponding to $0<u<5$ (a) and S2 corresponding to $-5<u<0$ (b). The only defective degeneracy when $g=0$ is EL($-$) which arises when $u=0$. The surfaces are spanned with $-5<\gamma_{-}<5$. (c) Parts of S1 (purple) and S2 (green) are achieved by setting $g=\pm \sqrt{u}/2 + \gamma_{-}/4$. Positive $u$ spans S1 and S2, while $u=0$ spans EL($-$). This surface is fully exceptional, except when $g$ is made zero via $\gamma_{-}=\pm 2\sqrt{u}$ (pink dashed line, DL3). Parameter ranges used for mapping are $0<u<5, \; -5<\gamma_{-}<5$. }
    \label{fig:S12}
\end{figure}
The swallowtail can be made both diabolical or exceptional depending on how the Hamiltonian parameters are tuned. For example, in our model with $\delta\omega_2=0$, the entire surface is diabolical when the two modes are decoupled, i.e., $g=0$  as depicted in Fig. \hyperref[fig:S12]{S3(a)-S3(b)}, with positive (negative) $u$ spanning subsurfaces S1 (S2). The determinant of the eigenvector matrix from Eq.~\eqref{eq::eig-vec-matr} is then equal to $2e^{-i\phi_{1}} \sqrt{u}/\xi_1$, which is only zero when $u=0$ (green line on Fig.~\ref{fig:S12}), corresponding to the defective EL($-$). However, parts of the swallowtail can also be achieved by setting $g=\pm \sqrt{u}/2+\gamma_{-}/4$. Then S2 (green on Fig. \hyperref[fig:S12]{S3(c)}) is fully exceptional, and S1 (purple on Fig. \hyperref[fig:S12]{S3(c)}) only becomes diabolical when the combination of $u, \; \gamma_{-}$ reaches zero at DL3, (pink dashed lines on Fig. \hyperref[fig:S12]{S3(c)}).
More specifically, S2 is spanned with 
\fla{
g=\sqrt{u}/2+\gamma_{-}, \qquad \gamma_{-}>0 \qquad 
\text{or} \qquad
g=-\sqrt{u}/2+\gamma_{-}, \qquad \gamma_{-}<0,}
and S1 is spanned if the conditions on $\gamma_{-}$ are switched. We note that in the main text we use $g=0$ mapping only for areas of the swallowtail that are not achievable otherwise.